\documentclass[RNAAS]{aastex62}

\begin{document}

\title{Gravitational microlensing events from the first year of the northern Galactic plane survey by the Zwicky Transient Facility}

\correspondingauthor{Przemek Mr\'oz}
\email{pmroz@astro.caltech.edu}

\author[0000-0001-7016-1692]{Przemek Mr\'oz}
\affiliation{Division of Physics, Mathematics, and Astronomy, California Institute of Technology, Pasadena, CA 91125, USA}
\author[0000-0001-6279-0552]{R.~A.~Street}
\affiliation{Las Cumbres Observatory Global Telescope Network, 6740 Cortona Drive, Suite 102, Goleta, CA 93117, USA}
\author[0000-0002-6578-5078]{E.~Bachelet}
\affiliation{Las Cumbres Observatory Global Telescope Network, 6740 Cortona Drive, Suite 102, Goleta, CA 93117, USA}
\author[0000-0002-6786-8774]{E.~O.~Ofek}
\affiliation{Department of Particle Physics and Astrophysics, Weizmann Institute of Science, 234 Herzl Street, 76100 Rehovot, Israel}
\author[0000-0001-8018-5348]{E.~C.~Bellm}
\affiliation{DIRAC Institute, Department of Astronomy, University of Washington, 3910 15th Avenue NE, Seattle, WA 98195, USA}
\author{R.~Dekany}
\affiliation{Caltech Optical Observatories, California Institute of Technology, Pasadena, CA 91125, USA}
\author[0000-0001-5060-8733]{D.~A.~Duev}
\affiliation{Division of Physics, Mathematics, and Astronomy, California Institute of Technology, Pasadena, CA 91125, USA}
\author[0000-0002-3653-5598]{A.~Gal-Yam}
\affiliation{Department of Particle Physics and Astrophysics, Weizmann Institute of Science, 234 Herzl Street, 76100 Rehovot, Israel}
\author[0000-0002-3168-0139]{M.~J.~Graham}
\affiliation{Division of Physics, Mathematics, and Astronomy, California Institute of Technology, Pasadena, CA 91125, USA}
\author[0000-0002-8532-9395]{F.~J.~Masci}
\affiliation{IPAC, California Institute of Technology, 1200 E. California
             Blvd, Pasadena, CA 91125, USA}
\author{M.~Porter}
\affiliation{Caltech Optical Observatories, California Institute of Technology, Pasadena, CA 91125, USA}
\author[0000-0001-7648-4142]{B.~Rusholme}
\affiliation{IPAC, California Institute of Technology, 1200 E. California
             Blvd, Pasadena, CA 91125, USA}
\author{R.~M.~Smith}
\affiliation{Caltech Optical Observatories, California Institute of Technology, Pasadena, CA 91125, USA}
\author[0000-0001-6753-1488]{M.~T.~Soumagnac}
\affiliation{Lawrence Berkeley National Laboratory, 1 Cyclotron Road, Berkeley, CA 94720, USA}
\affiliation{Department of Particle Physics and Astrophysics, Weizmann Institute of Science, 234 Herzl Street, 76100 Rehovot, Israel}
\author{J.~Zolkower}
\affiliation{Caltech Optical Observatories, California Institute of Technology, Pasadena, CA 91125, USA}

\keywords{gravitational lensing: micro}

\section{} 

The Zwicky Transient Facility (ZTF) \citep{bellm2019,graham2019,masci2019} is currently surveying the entire northern sky, including dense Galactic plane fields. Here, we present preliminary results of the search for gravitational microlensing events in the ZTF data collected from the beginning of the survey (March 20, 2018) through June 30, 2019.

Searches for gravitational microlensing events have been traditionally confined to the Galactic bulge, where the probability of microlensing (and the event rate) is the highest. However, in recent years a number of bright events were discovered outside the Galactic bulge. For example, \citet{nucita2018} found a super-Earth-mass planet in a high-magnification event TCP~J05074264+2447555 detected toward the Galactic anticenter ($l\approx 179^{\circ}$). This was the first event with the two images generated by gravitational microlensing resolved with the interferometric observations \citep{dong2019}. \citet{wyrzyk2019} was able to measure all orbital parameters of the binary lens in the spectacular event Gaia16aye ($l\approx 65^{\circ}$).

Gravitational microlensing enables one to find all types of ``dark objects'', including neutron stars, single and binary black holes. While photometric observations alone are usually insufficient to determine masses of lensing objects, the combination of ground-based photometry and precise astrometric \textit{Gaia} satellite observations will enable achieving that goal \citep[e.g.,][]{lu2016}. Microlensing events located in the Galactic disk have, on average, larger angular Einstein radii than Galactic bulge events \citep{sajadian2019}, so the astrometric signal is stronger. \textit{Gaia} performance in the dense Galactic bulge fields is suboptimal so it is important to identify as many microlensing events as possible in the less crowded Galactic disk fields.

ZTF conducts the Galactic Plane Survey with nightly observations of all visible fields in the region $|b|<7^{\circ}$, $\delta > -31^{\circ}$ in $g$ and $r$ bands. Additionally, Galactic plane fields are observed as part of ZTF collaboration and Caltech surveys. 

We plan to carry out a comprehensive analysis of microlensing events in the ZTF footprint (including the measurements of the microlensing optical depth and event rate) in the future. Here, we present the first discoveries based on the first $\sim15$ months of the survey that demonstrate that the current observing strategy enables the identification and characterization of microlensing events in the Galactic disk fields.

Our methodology is similar to that used in our previous works \citep{mroz2017,mroz2019}. We analyzed $r$-band light curves of objects associated with ZTF alerts in 408 fields at low Galactic latitudes ($|b|\leq 20^{\circ}$); we required at least five alert detections (meaning that the object was detected on a difference image produced by the \citet{zackay2016} algorithm). Then, we searched for objects with at least three consecutive data points that are at least $3\sigma_{\rm base}$ brighter than the baseline flux $F_{\rm base}$, where $F_{\rm base}$ and $\sigma_{\rm base}$ were calculated using data points outside a 160 day window centered on the event. We also fitted the microlensing point-source point-lens model to the light curves of all candidate objects. We selected candidate events that 1) do not exhibit any variability outside the window centered on the event, and 2) can be well-described by a microlensing point-lens point-source model. The light curves of selected candidates were additionally visually vetted by a human expert. The final models are based on simultaneous modeling of $g$- and $r$-band light curves.

\begin{figure}
\figurenum{1}
\includegraphics[width=0.49\textwidth]{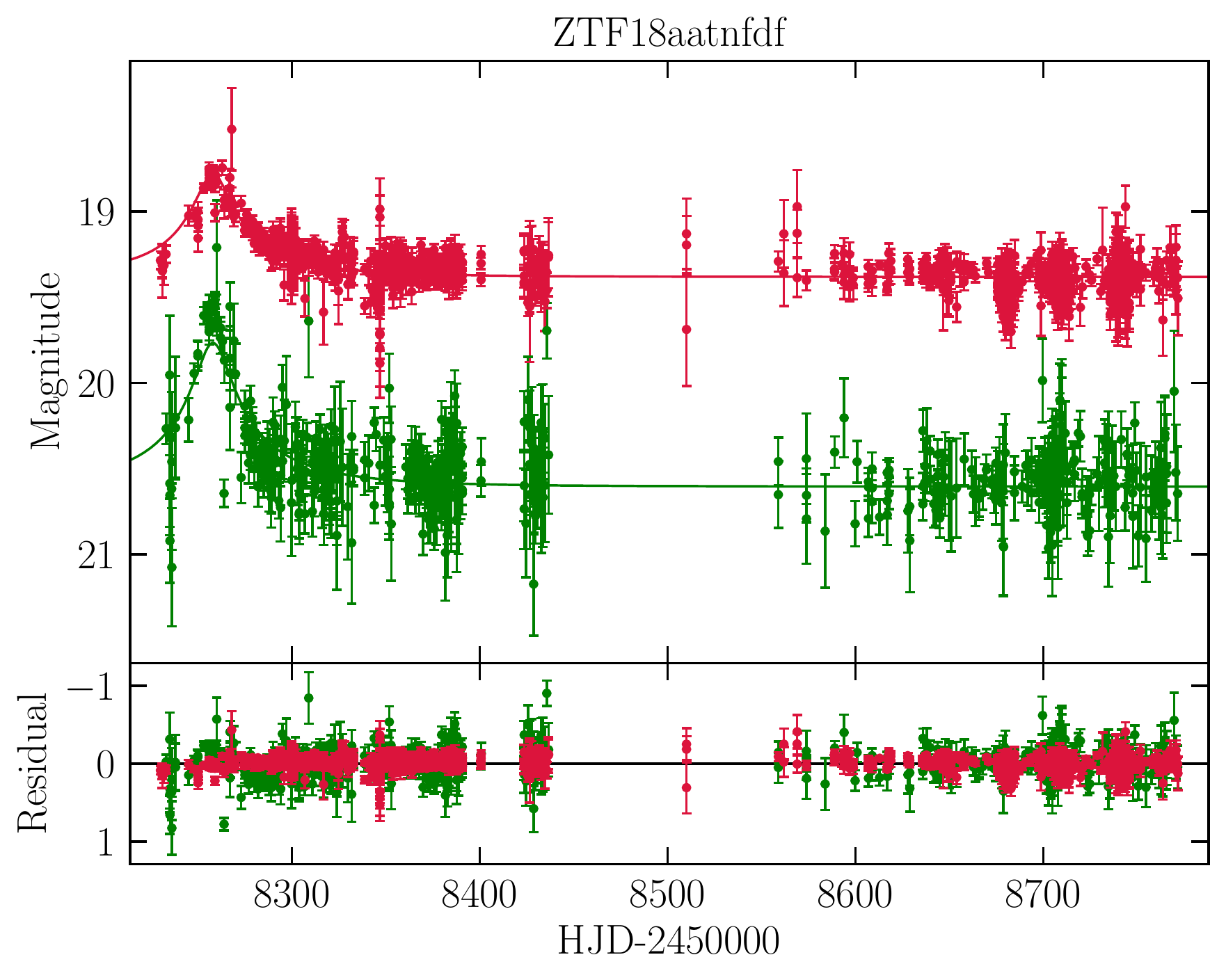}
\includegraphics[width=0.49\textwidth]{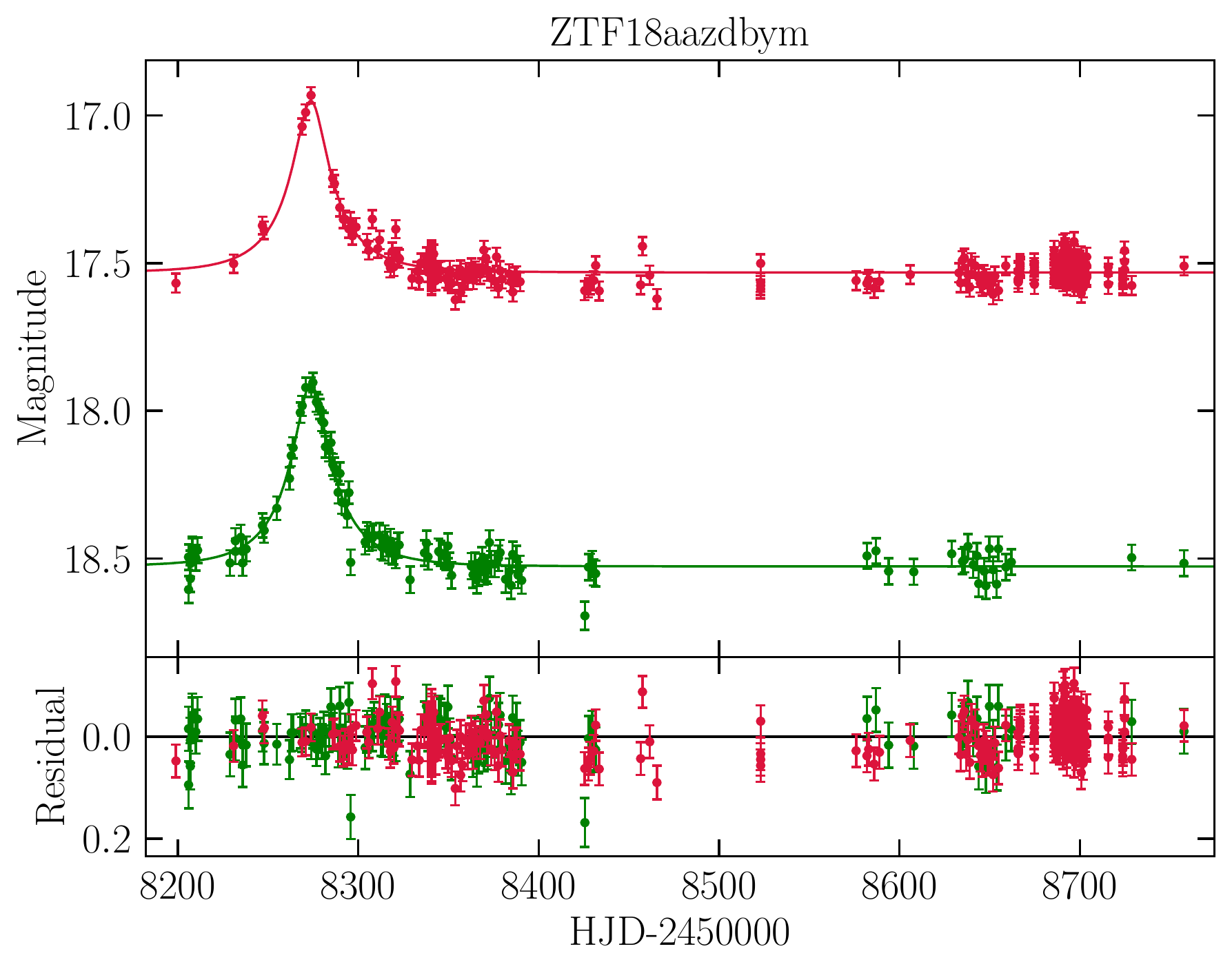}
\includegraphics[width=0.49\textwidth]{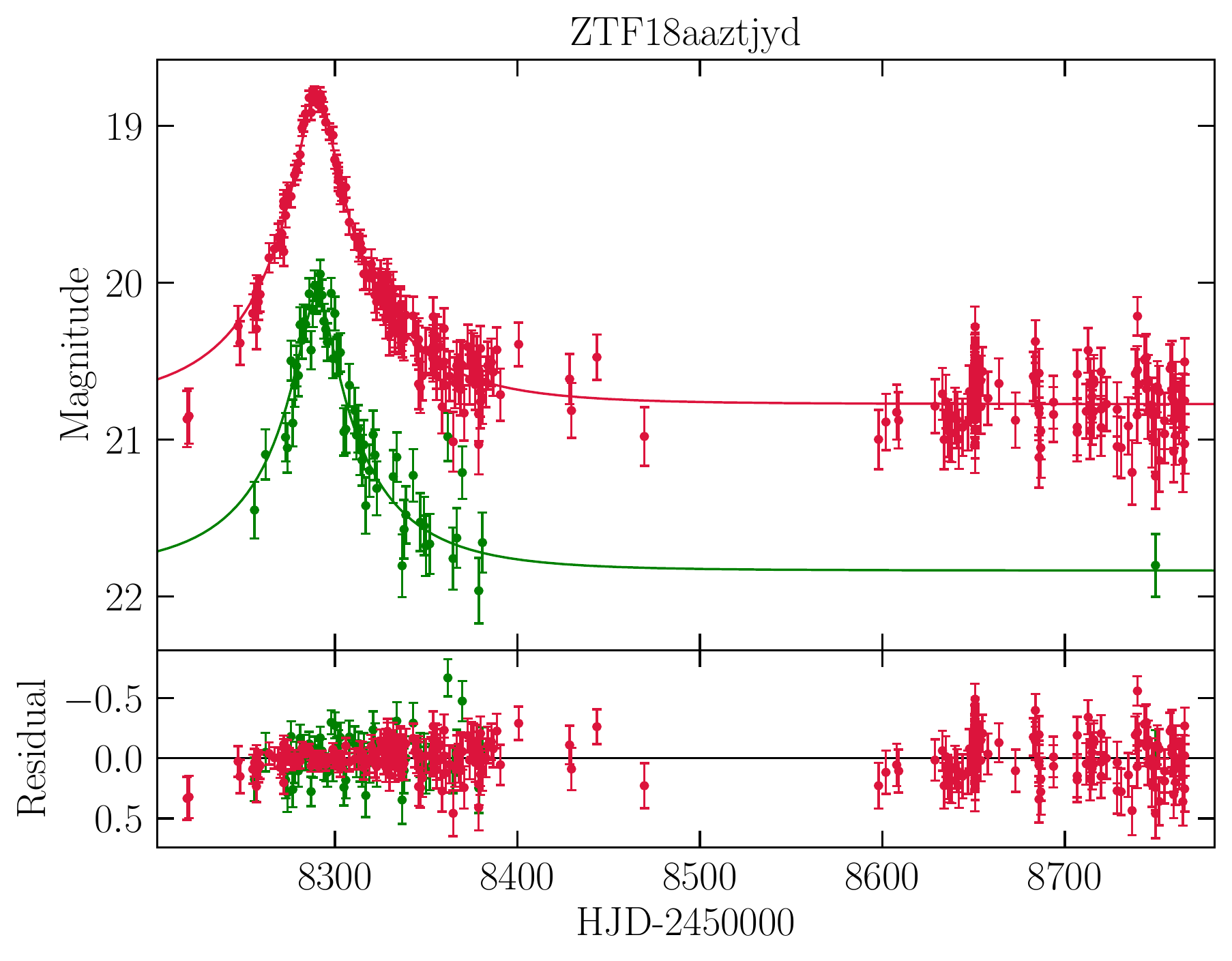}
\includegraphics[width=0.49\textwidth]{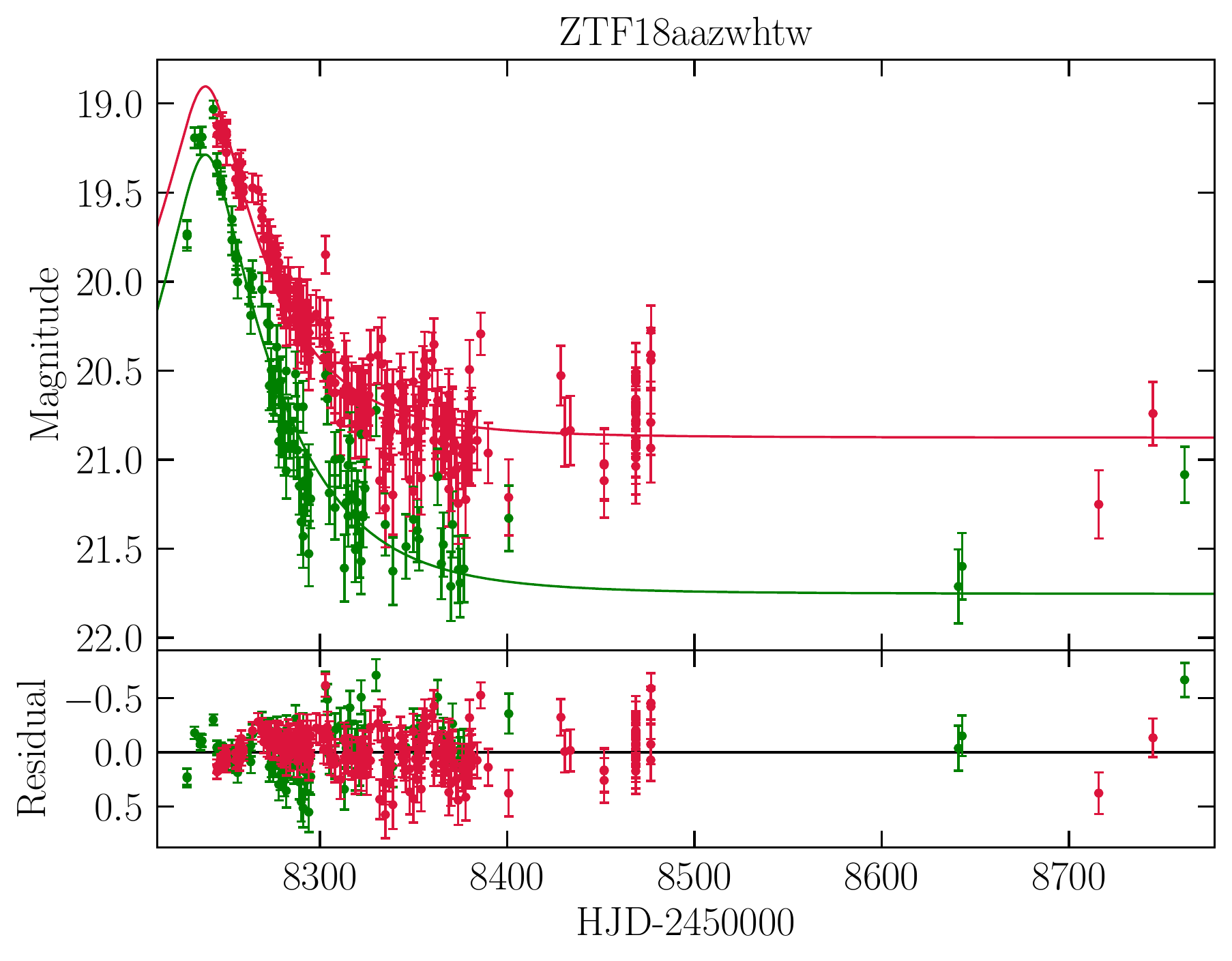}
\includegraphics[width=0.49\textwidth]{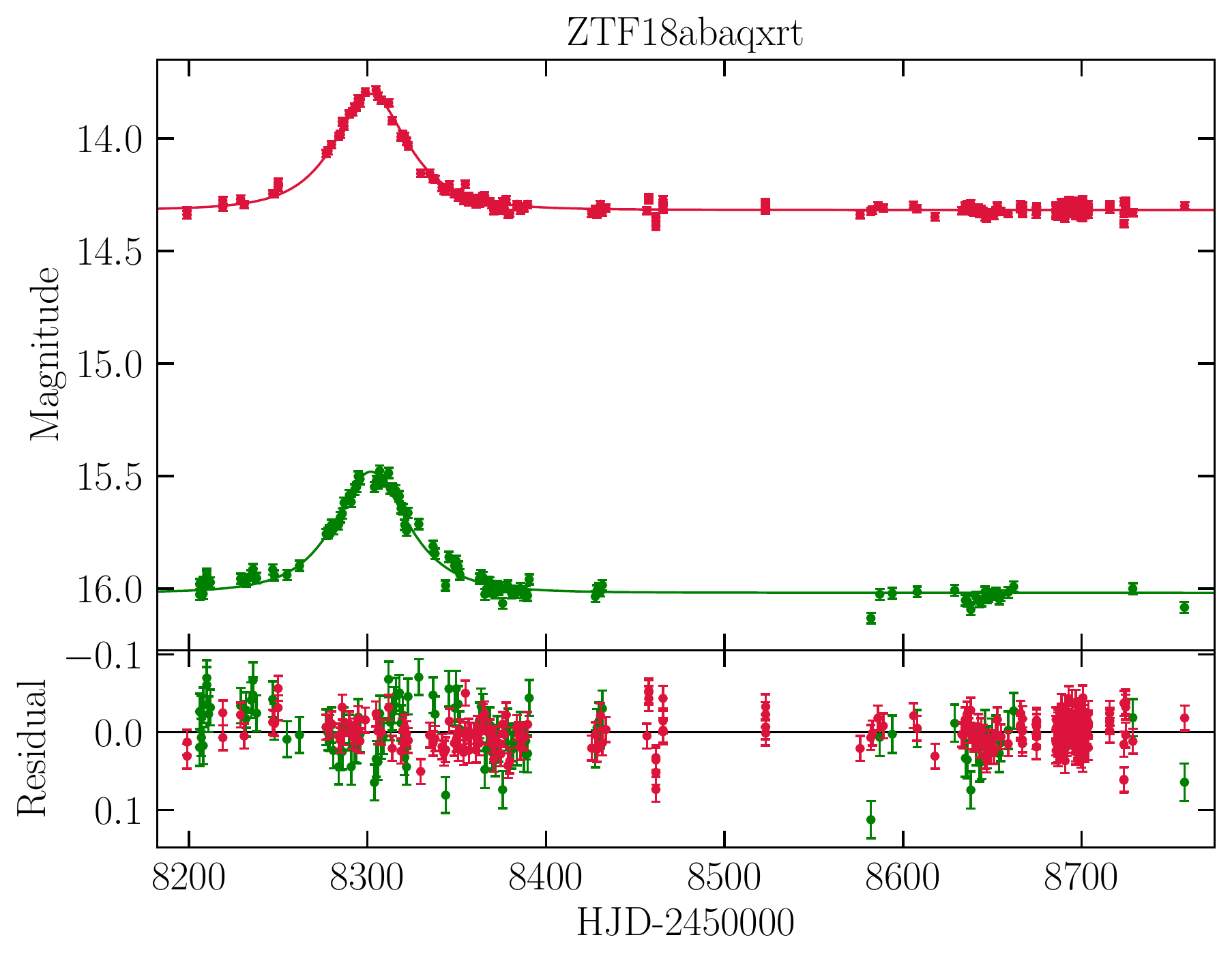}
\includegraphics[width=0.49\textwidth]{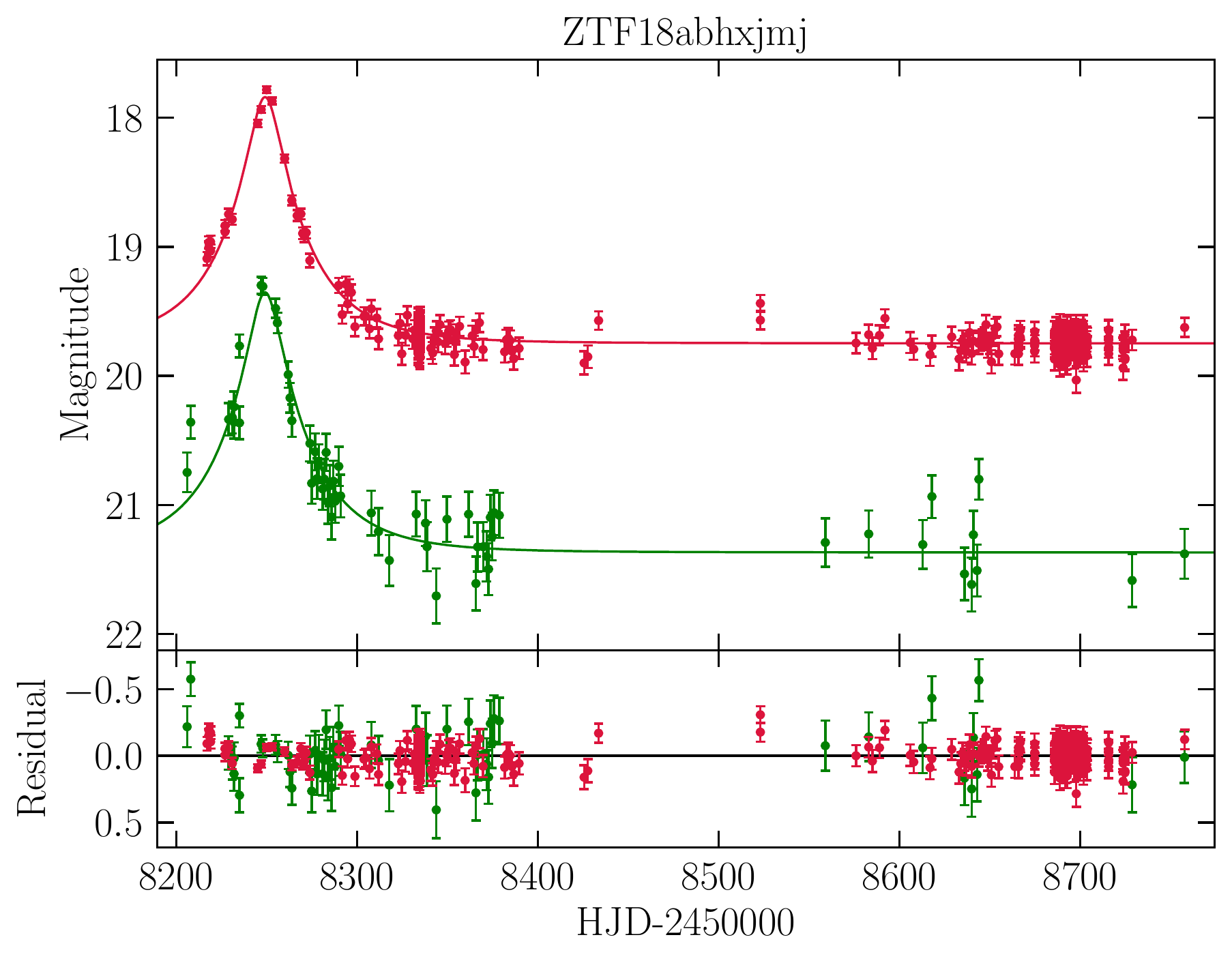}
\caption{Light curves of detected microlensing events. Green and red points were taken in $g$ and $r$ filters, respectively.}
\label{fig:lc}
\end{figure}

We found 30 likely events which are listed in Table~\ref{tab:events} and shown in Figure~\ref{fig:lc}. The  best-fitting model parameters are presented in Table~\ref{tab:models}. Although the current sample is relatively small, the properties of detected events are different from those of Galactic bulge events. All detected events have relatively long Einstein timescales ($30 \lesssim t_{\rm E} \lesssim 200$\,d) whereas typical timescales of bulge events are shorter ($t_{\rm E} \sim 20$\,d) \citep[e.g.,][]{mroz2017}. This may be partly explained by selection biases, but we have demonstrated \citep{mroz2019} that nightly observations are sufficient to detect events with timescales as short as a few days. From the theoretical point of view \citep[e.g.,][]{sajadian2019}, it is expected that, on average, Galactic plane events should be longer than those in the bulge direction because the source and the lens, both in the Galactic disk, are moving in a similar direction. 

For four of the detected events, we were able to measure microlens parallax. ZTF18abaqxrt is particularly interesting because the source is bright ($r\approx 14.7$) and so the \textit{Gaia} satellite have likely measured the astrometric microlensing signal. We also detected five likely binary microlensing events (ZTF18ablruzq, ZTF18abqawpf, ZTF18abqazwf, ZTF18acskgwu, ZTF19aatudnj). Preliminary models indicate that these events were caused by stellar binaries. 

\acknowledgments

We thank Jan Skowron for providing software for computation of microlensing magnifications. Based on observations obtained with the Samuel Oschin Telescope 48-inch and the 60-inch Telescope at the Palomar Observatory as part of the Zwicky Transient Facility project. ZTF is supported by the National Science Foundation under Grant No. AST-1440341 and a collaboration including Caltech, IPAC, the Weizmann Institute for Science, the Oskar Klein Center at Stockholm University, the University of Maryland, the University of Washington, Deutsches Elektronen-Synchrotron and Humboldt University, Los Alamos National Laboratories, the TANGO Consortium of Taiwan, the University of Wisconsin at Milwaukee, and Lawrence Berkeley National Laboratories. Operations are conducted by COO, IPAC, and UW.

\begin{deluxetable}{lrrrrl}
\tablecaption{Gravitational microlensing events in the ZTF DR2 data}
\tablehead{
\colhead{Event} & \colhead{R.A.} & \colhead{Decl.} & \colhead{$l$} & \colhead{$b$} & \colhead{Remarks}
}
\startdata
ZTF18aatnfdf        & 286.633211 &  32.248996 &  63.592037 &  11.173057 &  \\
ZTF18aazdbym        & 290.784286 &   7.810517 &  43.509003 &  --3.416870 &  \\
ZTF18aaztjyd        & 326.173116 &  59.377872 & 101.101575 &   4.669597 &  \\
ZTF18aazwhtw        & 339.955528 &  51.647223 & 103.116644 &  --6.088552 &  \\
ZTF18abaqxrt        & 290.617225 &   1.706486 &  38.010990 &  --6.113752 &  \\
ZTF18abhxjmj        & 284.029167 &  13.152260 &  45.192580 &   4.937164 &  \\
ZTF18ablrbkj        & 271.850400 & --10.314477 &  18.695441 &   4.908538 &  \\
ZTF18ablrdcc        & 271.439120 & --12.014556 &  17.006029 &   4.441709 & Gaia18chq \\
ZTF18ablruzq         & 284.338291 &  11.433438 &  43.790788 &   3.892060 & binary \\
ZTF18abmoxlq        & 285.984027 & --13.929477 &  21.832965 &  --9.024192 &  \\
ZTF18abnbmsr        & 307.149376 &  22.830478 &  64.600302 &  --9.235573 & Gaia18cmk \\
ZTF18abqawpf         & 287.113964 &   1.531903 &  36.239259 &  --3.084660 & binary \\
ZTF18abqazwf\tablenotemark{a}         & 285.134471 &  30.511120 &  61.434070 &  11.599466 & binary? \\
ZTF18abqbeqv        & 279.578723 &   7.837854 &  38.448164 &   6.467501 &  \\
ZTF18absrqlr        & 307.149376 &  22.830478 &  64.600302 &  --9.235573 &  \\
ZTF18abtnvsg        & 291.019150 &  20.478976 &  54.790638 &   2.361105 &  \\
ZTF18acskgwu         &  76.632447 &   8.425664 & 192.546959 & --18.799234 & binary \\
ZTF19aabbuqn        &  48.694244 &  62.343390 & 138.738918 &   3.955177 &  \\
ZTF19aaekacq        & 279.404621 &  11.200516 &  41.407932 &   8.116296 &  \\
ZTF19aainwvb        &  55.197569 &  57.955805 & 143.884446 &   2.147052 & Gaia19bjq \\
ZTF19aamlgyh        & 289.114418 &  26.653532 &  59.467936 &   6.770307 & Gaia19asx \\
ZTF19aamrjmu        & 280.734529 &  32.873054 &  62.121495 &  15.974788 &  \\
ZTF19aaonska        & 273.900566 &  --2.256985 &  26.802123 &   6.922664 & Gaia19awc \\
ZTF19aaprbng        & 274.913476 &   0.590991 &  29.819420 &   7.338351 &  \\
ZTF19aatudnj\tablenotemark{a}         & 290.663294 &  19.550373 &  53.813242 &   2.218551 & Gaia19bzf, binary? \\
ZTF19aatwaux        & 258.208411 & --27.182057 & 357.476262 &   7.002211 &  \\
ZTF19aavisrq        & 297.706148 &  34.637344 &  70.054958 &   4.102724 & Gaia19dae \\
ZTF19aavndrc        & 281.836951 &  --4.338099 &  28.604991 &  --1.068892 &  \\
ZTF19aavnrqt        & 309.034132 &  32.720880 &  73.669581 &  --4.807238 &  \\
ZTF19aaxsdqz        & 283.497170 &  --1.152267 &  32.197043 &  --1.092856 &  \\
\enddata
\tablenotetext{a}{Possible microlensing event.}
\tablecomments{Equatorial coordinates are given for the epoch J2000.}
\label{tab:events}
\end{deluxetable}

\begin{deluxetable}{lrrrrrrrrrr}
\tablecaption{Best-fitting parameters of point-lens point-source events}
\tablehead{
\colhead{Event} & \colhead{$t_{\rm 0,par}$} & \colhead{$t_0$} & \colhead{$t_{\rm E}$} & \colhead{$u_0$} & \colhead{$\pi_{\mathrm{E},N}$} & \colhead{$\pi_{\mathrm{E},E}$} & \colhead{$r_{\rm s}$} & \colhead{$f_{\rm s}$} & \colhead{$\chi^2/\mathrm{dof}$} \\
& (HJD$'$) & (HJD$'$) & (d) & & & & (mag) & & & 
}
\startdata
ZTF18aatnfdf    & 8254 & $8257.86^{+0.32}_{-0.30}$ & $ 76.30^{+7.31}_{-6.69}$ & $ 0.128^{+0.017}_{-0.015}$ & & & $21.81^{+0.14}_{-0.15}$ & $ 0.11^{+0.02}_{-0.01}$ & $3388.9/1788$ & \\
ZTF18aazdbym    & 8273 & $8273.77^{+0.28}_{-0.28}$ & $ 32.16^{+4.54}_{-4.25}$ & $ 0.273^{+0.068}_{-0.050}$ & & & $19.01^{+0.26}_{-0.31}$ & $ 0.26^{+0.08}_{-0.06}$ & $494.8/492$ & \\
ZTF18aaztjyd    & 8290 & $8290.11^{+0.19}_{-0.20}$ & $ 68.85^{+7.84}_{-6.74}$ & $ 0.132^{+0.019}_{-0.017}$ & & & $21.04^{+0.16}_{-0.16}$ & $ 0.78^{+0.12}_{-0.10}$ & $405.4/361$ & \\
ZTF18aazwhtw    & 8235 & $8238.84^{+0.46}_{-0.49}$ & $ 63.04^{+7.16}_{-5.78}$ & $ 0.228^{+0.037}_{-0.033}$ & & & $20.44^{+0.20}_{-0.19}$ & $ 1.51^{+0.27}_{-0.23}$ & $589.0/318$ & \\
ZTF18abaqxrt    & 8301 & $8301.93^{+0.21}_{-0.20}$ & $ 30.32^{+2.11}_{-1.32}$ & $ 0.673^{+0.050}_{-0.068}$ & & & $14.51^{+0.19}_{-0.14}$ & $ 0.84^{+0.11}_{-0.14}$ & $683.1/387$ & \\
ZTF18abaqxrt    & 8301 & $8302.75^{+0.24}_{-0.23}$ & $ 33.91^{+4.15}_{-3.35}$ & $-0.581^{+0.096}_{-0.103}$ & $-0.83^{+0.15}_{-0.15}$ & $-0.00^{+0.28}_{-0.28}$ & $14.78^{+0.30}_{-0.30}$ & $ 0.65^{+0.20}_{-0.16}$ & $645.5/385$ & \\
ZTF18abaqxrt    & 8301 & $8302.86^{+0.28}_{-0.28}$ & $ 33.26^{+4.40}_{-2.98}$ & $ 0.608^{+0.089}_{-0.092}$ & $-1.42^{+0.25}_{-0.29}$ & $-0.30^{+0.27}_{-0.23}$ & $14.69^{+0.28}_{-0.25}$ & $ 0.71^{+0.18}_{-0.16}$ & $645.3/385$ & \\
ZTF18abhxjmj    & 8249 & $8249.14^{+0.21}_{-0.21}$ & $ 46.04^{+2.58}_{-2.35}$ & $ 0.183^{+0.016}_{-0.015}$ & & & $19.68^{+0.10}_{-0.10}$ & $ 1.06^{+0.10}_{-0.09}$ & $539.2/450$ & \\
ZTF18abhxjmj    & 8249 & $8250.50^{+0.23}_{-0.22}$ & $ 75.08^{+11.64}_{-8.82}$ & $ 0.088^{+0.016}_{-0.014}$ & $ 0.16^{+0.11}_{-0.47}$ & $-0.72^{+0.21}_{-0.09}$ & $20.53^{+0.19}_{-0.19}$ & $ 0.49^{+0.09}_{-0.08}$ & $450.9/448$ & \\
ZTF18abhxjmj    & 8249 & $8250.65^{+0.23}_{-0.24}$ & $ 79.85^{+11.05}_{-8.68}$ & $-0.087^{+0.013}_{-0.014}$ & $ 0.12^{+0.09}_{-0.30}$ & $-0.71^{+0.15}_{-0.09}$ & $20.54^{+0.18}_{-0.17}$ & $ 0.48^{+0.08}_{-0.07}$ & $451.8/448$ & \\
ZTF18ablrbkj    & 8261 & $8260.87^{+0.99}_{-1.13}$ & $171.57^{+228.23}_{-74.59}$ & $ 0.077^{+0.074}_{-0.047}$ & & & $22.43^{+1.04}_{-0.79}$ & $ 0.24^{+0.26}_{-0.15}$ & $189.0/176$ & \\
ZTF18ablrdcc    & 8353 & $8353.71^{+0.39}_{-0.42}$ & $ 47.76^{+4.93}_{-4.06}$ & $ 0.190^{+0.028}_{-0.028}$ & & & $19.96^{+0.17}_{-0.15}$ & $ 1.25^{+0.19}_{-0.18}$ & $278.3/266$ & \\
ZTF18abmoxlq    & 8323 & $8323.08^{+0.32}_{-0.30}$ & $ 51.40^{+4.65}_{-2.49}$ & $ 0.448^{+0.033}_{-0.052}$ & & & $17.81^{+0.20}_{-0.12}$ & $ 0.89^{+0.10}_{-0.15}$ & $489.4/598$ & \\
ZTF18abnbmsr    & 8363 & $8363.10^{+0.19}_{-0.18}$ & $ 38.20^{+3.07}_{-2.89}$ & $ 0.302^{+0.039}_{-0.032}$ & & & $18.97^{+0.15}_{-0.17}$ & $ 0.59^{+0.10}_{-0.08}$ & $707.6/537$ & \\
ZTF18abqbeqv    & 8383 & $8382.46^{+0.89}_{-0.89}$ & $148.18^{+61.34}_{-40.00}$ & $ 0.220^{+0.099}_{-0.070}$ & $-0.28^{+0.08}_{-0.11}$ & $-0.04^{+0.01}_{-0.02}$ & $19.49^{+0.46}_{-0.49}$ & $ 0.07^{+0.04}_{-0.03}$ & $573.0/558$ & \\
ZTF18abqbeqv    & 8383 & $8382.62^{+2.18}_{-2.03}$ & $174.92^{+87.56}_{-50.89}$ & $-0.301^{+0.114}_{-0.179}$ & $-0.17^{+0.06}_{-0.07}$ & $-0.04^{+0.02}_{-0.02}$ & $19.04^{+0.61}_{-0.69}$ & $ 0.11^{+0.10}_{-0.05}$ & $577.3/558$ & \\
ZTF18abqbeqv    & 8383 & $8383.42^{+1.47}_{-1.39}$ & $202.51^{+251.02}_{-72.24}$ & $ 0.176^{+0.133}_{-0.105}$ & & & $19.81^{+1.06}_{-0.74}$ & $ 0.05^{+0.05}_{-0.03}$ & $611.2/560$ & \\
ZTF18absrqlr    & 8363 & $8362.87^{+0.16}_{-0.16}$ & $ 37.53^{+2.89}_{-2.77}$ & $ 0.310^{+0.039}_{-0.032}$ & & & $18.93^{+0.15}_{-0.17}$ & $ 0.60^{+0.10}_{-0.08}$ & $648.2/490$ & \\
ZTF18abtnvsg    & 8220 & $8219.98^{+0.27}_{-0.32}$ & $108.56^{+38.17}_{-23.98}$ & $ 0.046^{+0.017}_{-0.014}$ & & & $21.93^{+0.39}_{-0.35}$ & $ 0.09^{+0.03}_{-0.03}$ & $430.5/287$ & \\
ZTF19aabbuqn    & 8506 & $8506.14^{+0.42}_{-0.41}$ & $ 25.80^{+5.29}_{-3.99}$ & $ 0.478^{+0.139}_{-0.121}$ & & & $19.03^{+0.45}_{-0.44}$ & $ 0.54^{+0.27}_{-0.18}$ & $763.0/296$ & \\
ZTF19aaekacq    & 8544 & $8544.72^{+0.54}_{-0.52}$ & $ 79.53^{+10.20}_{-9.20}$ & $ 0.224^{+0.076}_{-0.060}$ & & & $18.65^{+0.26}_{-0.28}$ & $ 0.21^{+0.06}_{-0.04}$ & $465.4/546$ & \\
ZTF19aainwvb    & 8656 & $8615.99^{+0.59}_{-0.60}$ & $137.66^{+6.89}_{-5.42}$ & $-0.098^{+0.029}_{-0.022}$ & $-0.25^{+0.01}_{-0.01}$ & $ 0.24^{+0.02}_{-0.02}$ & $18.48^{+0.09}_{-0.06}$ & $ 0.99^{+0.06}_{-0.08}$ & $1028.2/554$ & \\
ZTF19aainwvb    & 8656 & $8651.56^{+0.44}_{-0.45}$ & $134.57^{+2.67}_{-2.49}$ & $ 0.349^{+0.009}_{-0.009}$ & & & $18.21^{+0.04}_{-0.04}$ & $ 1.34^{+0.05}_{-0.04}$ & $2294.7/558$ & \\
ZTF19aamlgyh    & 8567 & $8567.08^{+0.16}_{-0.16}$ & $ 33.43^{+2.90}_{-2.43}$ & $ 0.149^{+0.018}_{-0.017}$ & & & $19.04^{+0.14}_{-0.13}$ & $ 0.84^{+0.11}_{-0.10}$ & $937.4/652$ & \\
ZTF19aamrjmu    & 8579 & $8579.78^{+0.11}_{-0.11}$ & $ 60.30^{+3.24}_{-3.00}$ & $ 0.148^{+0.011}_{-0.011}$ & & & $20.10^{+0.09}_{-0.08}$ & $ 0.78^{+0.06}_{-0.06}$ & $1875.8/1203$ & \\
ZTF19aaonska    & 8613 & $8613.12^{+0.34}_{-0.32}$ & $ 70.20^{+4.24}_{-3.83}$ & $ 0.280^{+0.025}_{-0.025}$ & & & $19.10^{+0.12}_{-0.11}$ & $ 0.89^{+0.09}_{-0.09}$ & $603.6/404$ & \\
ZTF19aaprbng    & 8629 & $8630.34^{+1.25}_{-1.22}$ & $112.61^{+31.69}_{-17.91}$ & $ 0.593^{+0.163}_{-0.170}$ & & & $19.13^{+0.56}_{-0.45}$ & $ 0.57^{+0.30}_{-0.23}$ & $515.6/469$ & \\
ZTF19aatwaux    & 8636 & $8636.95^{+0.52}_{-0.52}$ & $ 53.34^{+8.14}_{-5.64}$ & $ 0.162^{+0.040}_{-0.039}$ & & & $19.28^{+0.26}_{-0.22}$ & $ 0.79^{+0.18}_{-0.17}$ & $342.9/380$ & \\
ZTF19aavisrq    & 8651 & $8651.79^{+0.02}_{-0.02}$ & $ 96.39^{+6.61}_{-5.83}$ & $ 0.017^{+0.001}_{-0.001}$ & & & $20.58^{+0.08}_{-0.07}$ & $ 0.14^{+0.01}_{-0.01}$ & $564.3/595$ & \\
ZTF19aavndrc    & 8637 & $8637.42^{+0.15}_{-0.16}$ & $ 73.95^{+4.39}_{-3.91}$ & $ 0.095^{+0.008}_{-0.008}$ & & & $19.90^{+0.09}_{-0.09}$ & $ 0.39^{+0.03}_{-0.03}$ & $429.2/458$ & \\
ZTF19aavnrqt    & 8721 & $8721.03^{+0.53}_{-0.49}$ & $ 77.21^{+4.10}_{-2.25}$ & $ 0.631^{+0.027}_{-0.048}$ & & & $17.17^{+0.14}_{-0.08}$ & $ 0.92^{+0.07}_{-0.11}$ & $1651.8/673$ & \\
ZTF19aaxsdqz    & 8676 & $8676.54^{+0.14}_{-0.14}$ & $ 62.54^{+3.75}_{-2.90}$ & $ 0.180^{+0.013}_{-0.014}$ & & & $18.99^{+0.10}_{-0.08}$ & $ 1.01^{+0.08}_{-0.09}$ & $331.2/282$ & \\
\enddata
\tablecomments{HJD$'$=HJD$-2450000$. Microlensing point-source point-lens models have the following parameters: $t_0$ and $u_0$ -- time and separation during the closest approach between the lens and the source, $t_{\rm E}$ -- Einstein timescale, $\pi_{\mathrm{E},N}$ and $\pi_{\mathrm{E},E}$ -- North and East components of the microlensing parallax vector, $r_{\rm s}$ -- $r$-band brightness of the source, $f_{\rm s}$ -- blending parameter. $t_{\rm 0,par}$ defines the coordinate system for parallax measurements and is not a fit parameter.}
\label{tab:models}
\end{deluxetable} 

\clearpage

\begin{figure}
\figurenum{1}
\includegraphics[width=0.49\textwidth]{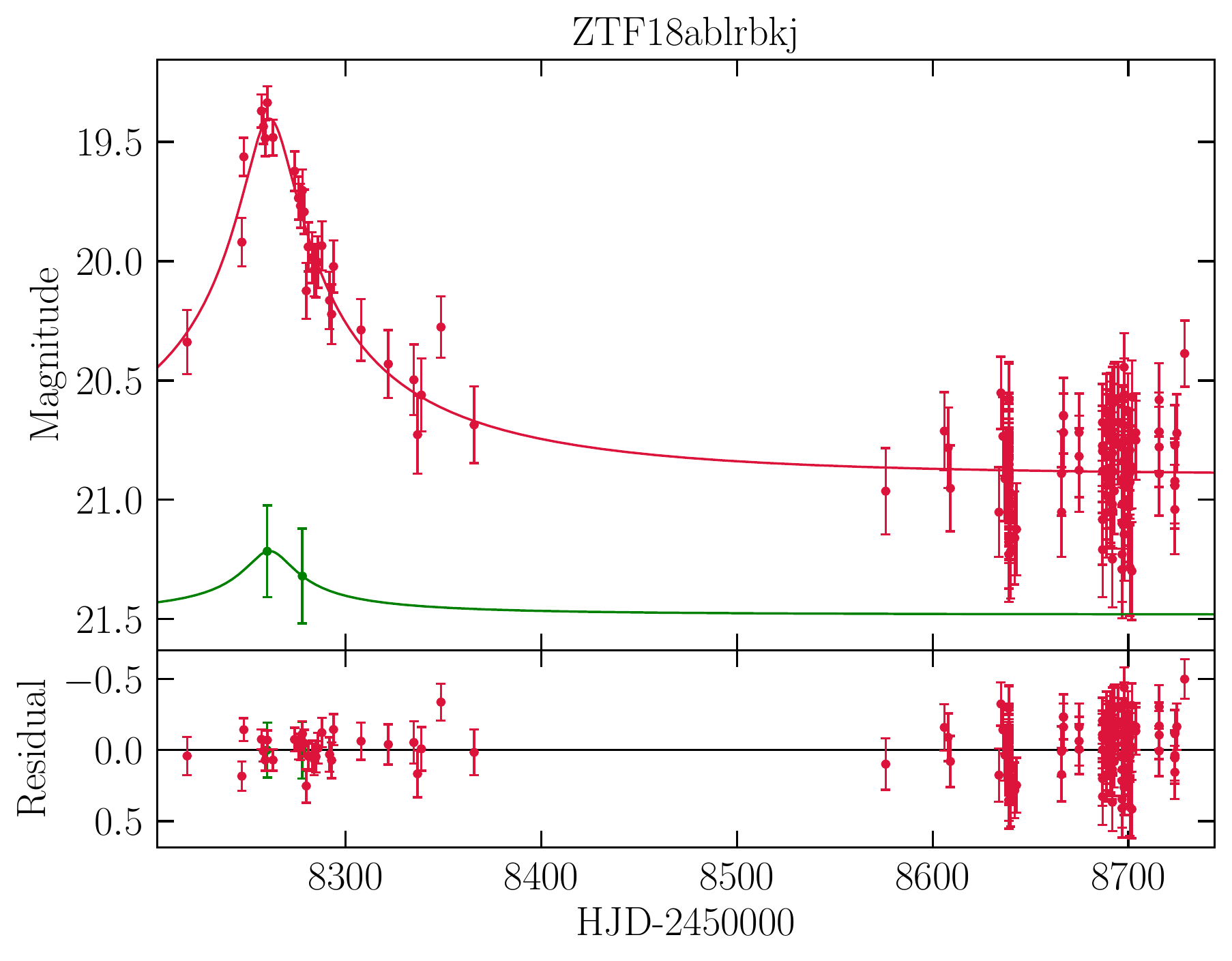}
\includegraphics[width=0.49\textwidth]{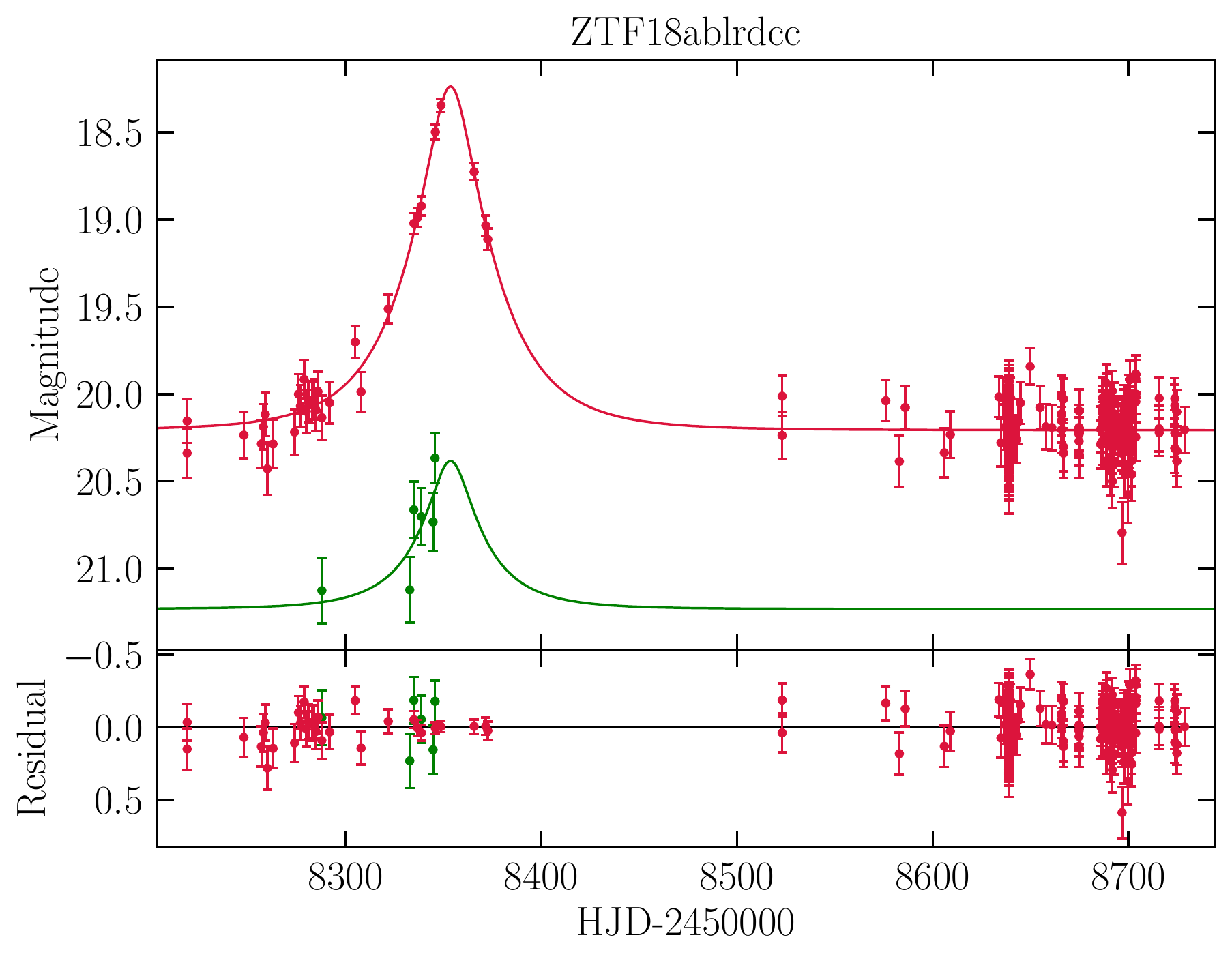}
\includegraphics[width=0.49\textwidth]{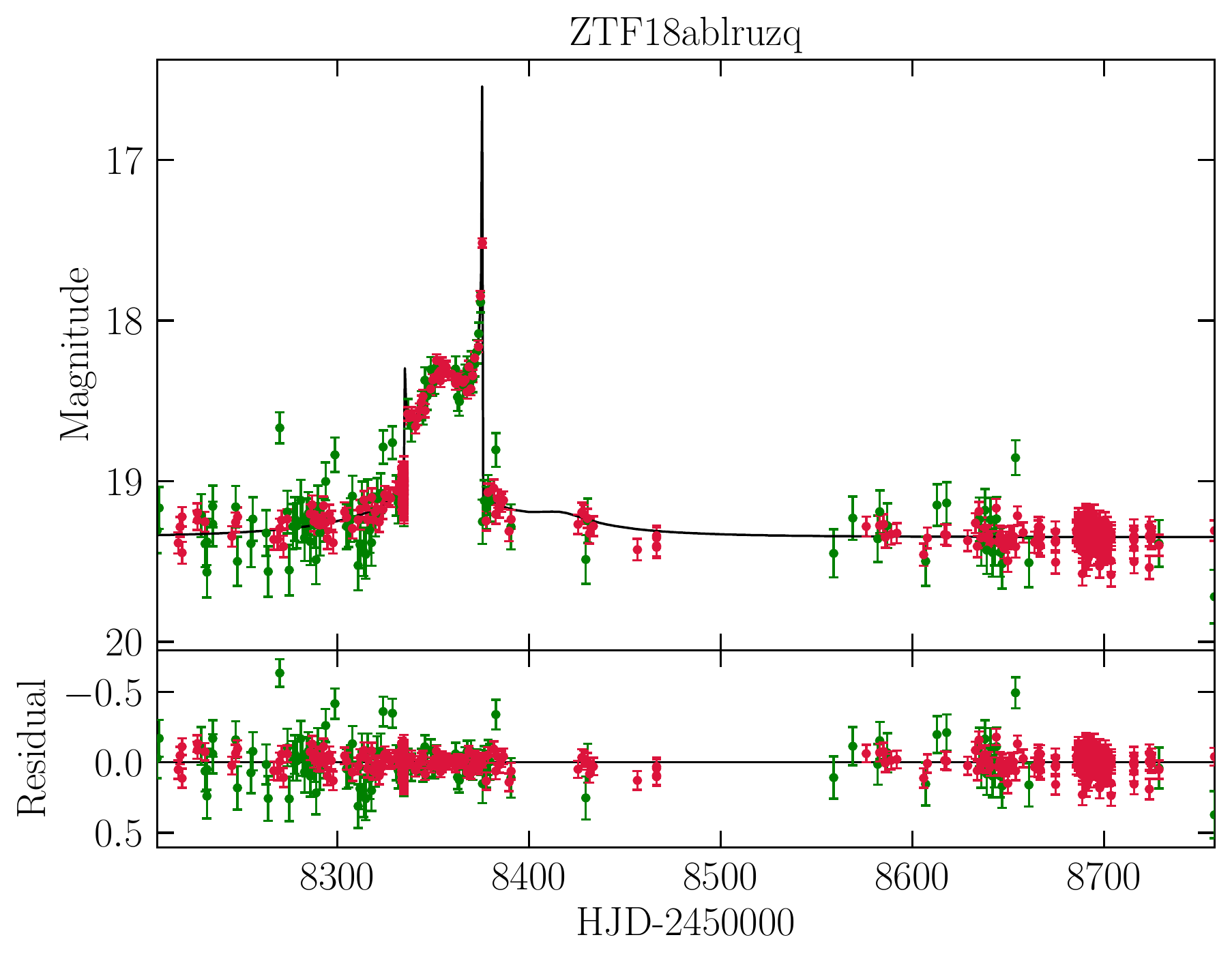}
\includegraphics[width=0.49\textwidth]{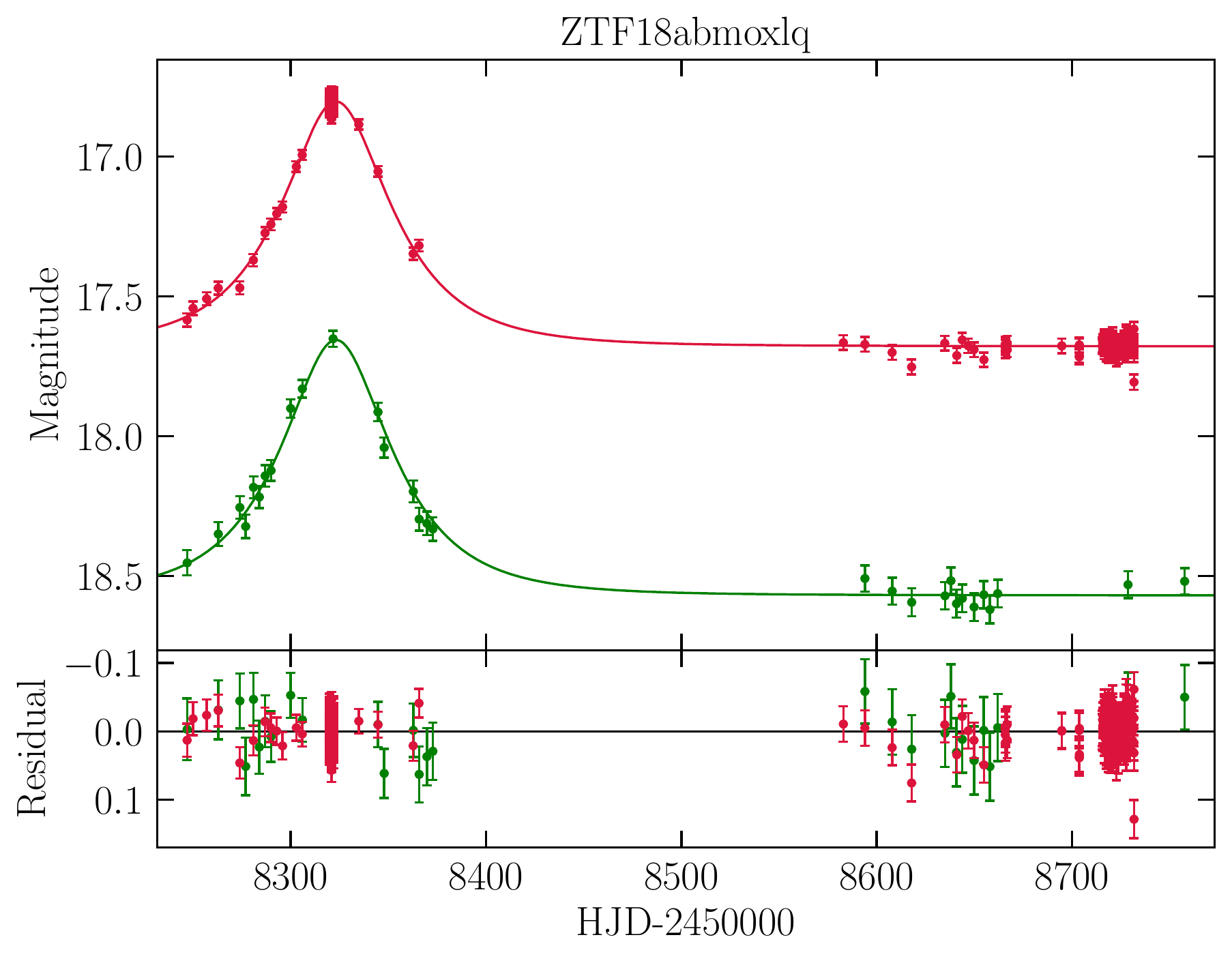}
\includegraphics[width=0.49\textwidth]{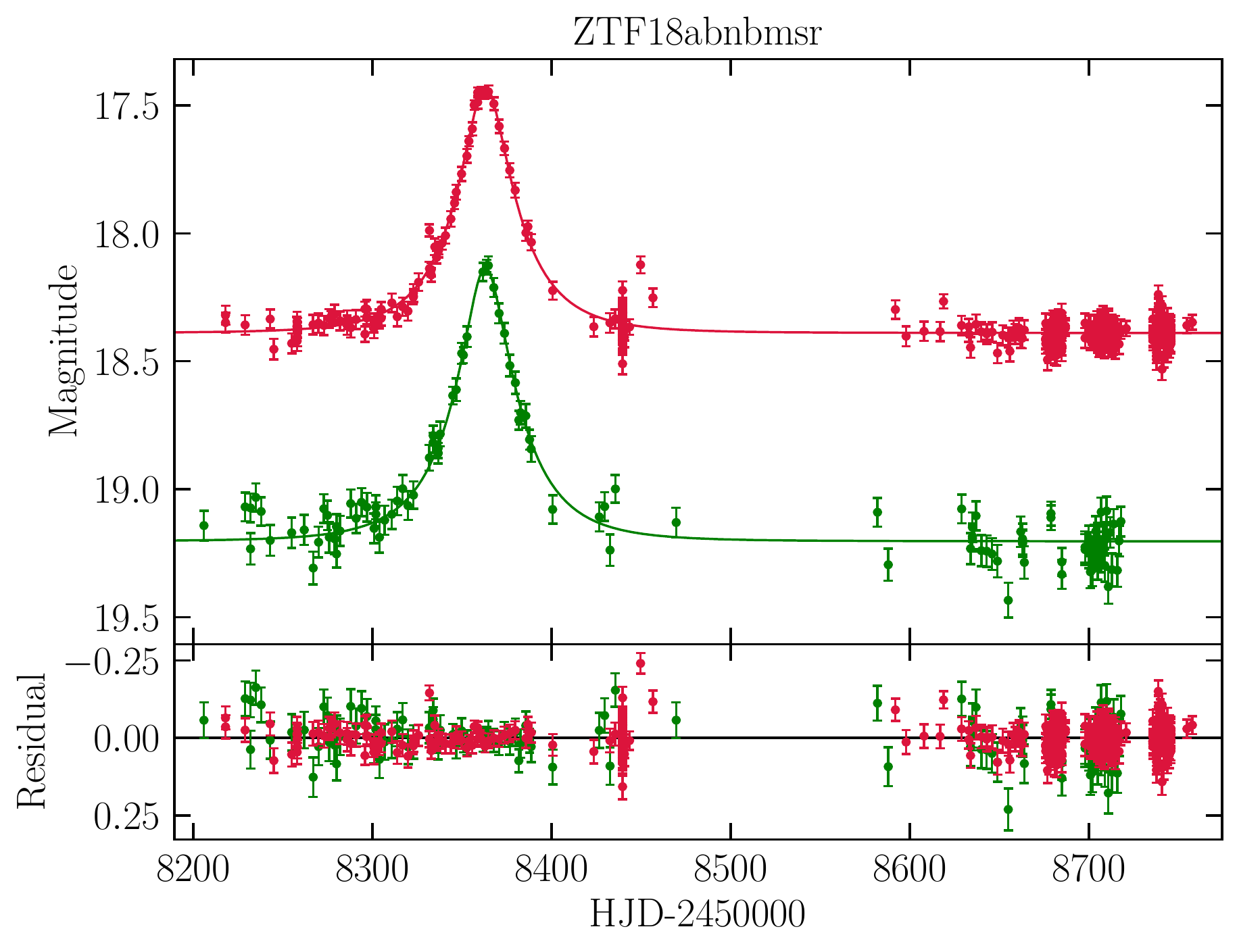}
\includegraphics[width=0.49\textwidth]{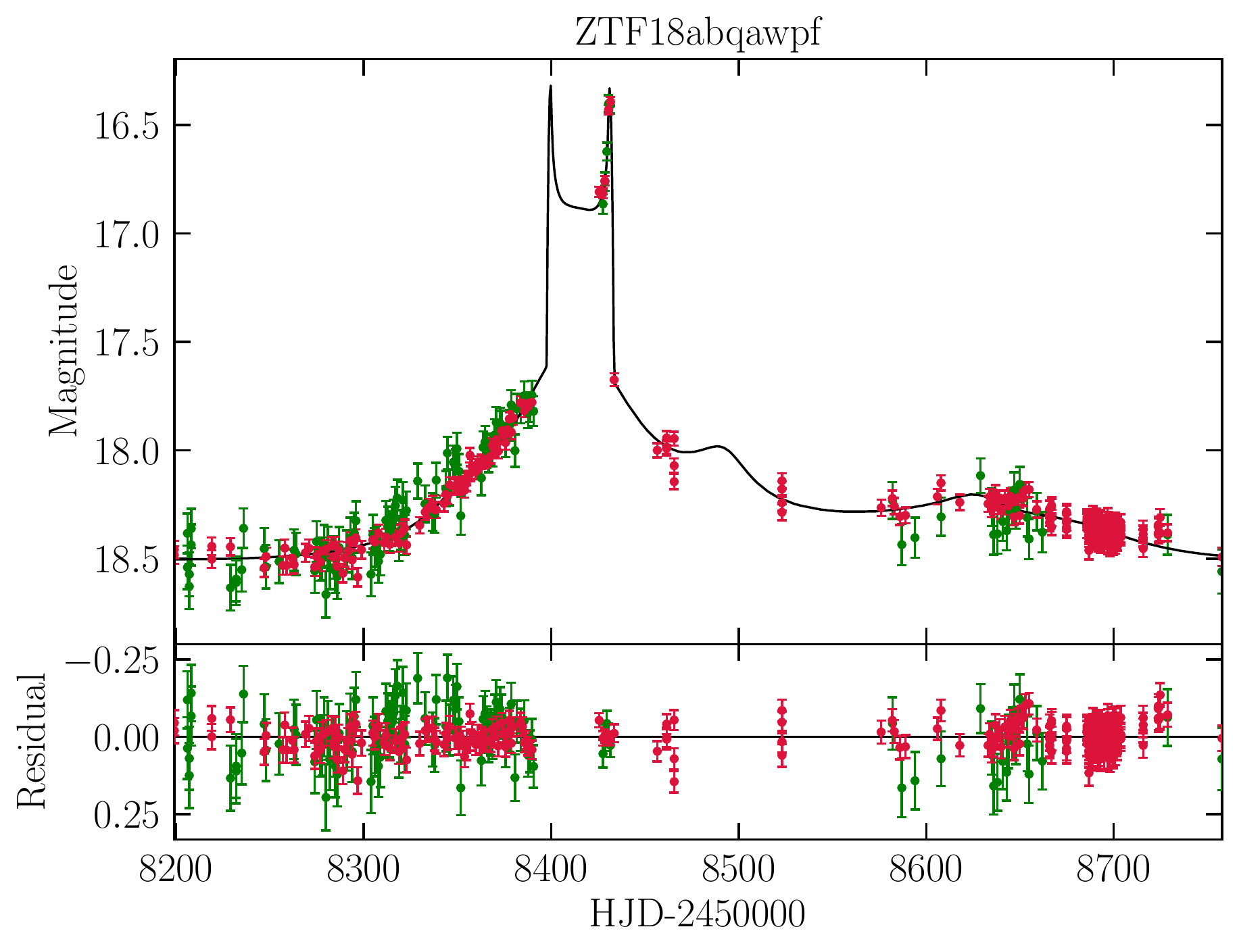}
\caption{Light curves of detected microlensing events. Green and red points were taken in $g$ and $r$ filters, respectively.}
\end{figure}

\begin{figure}
\figurenum{1}
\includegraphics[width=0.49\textwidth]{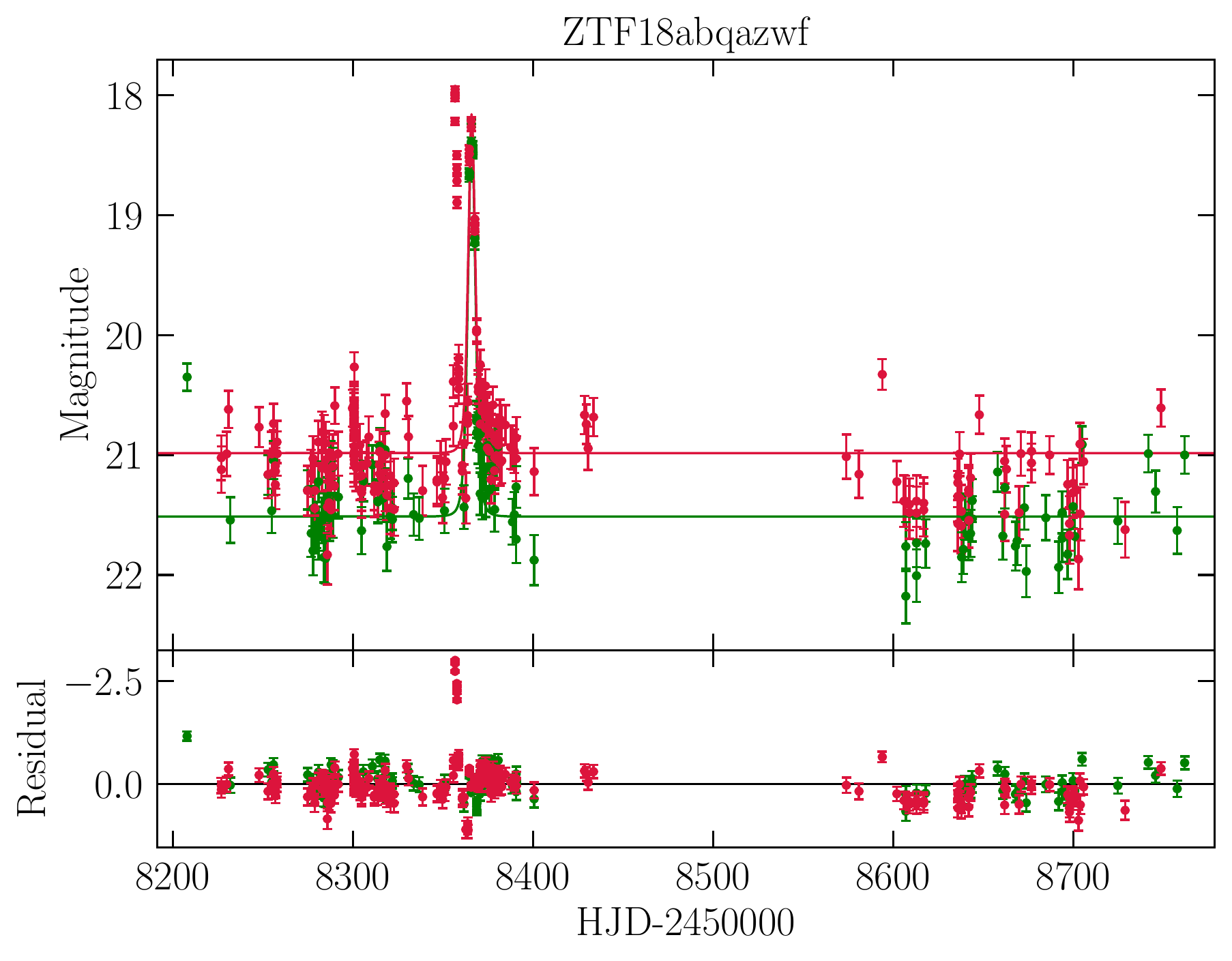}
\includegraphics[width=0.49\textwidth]{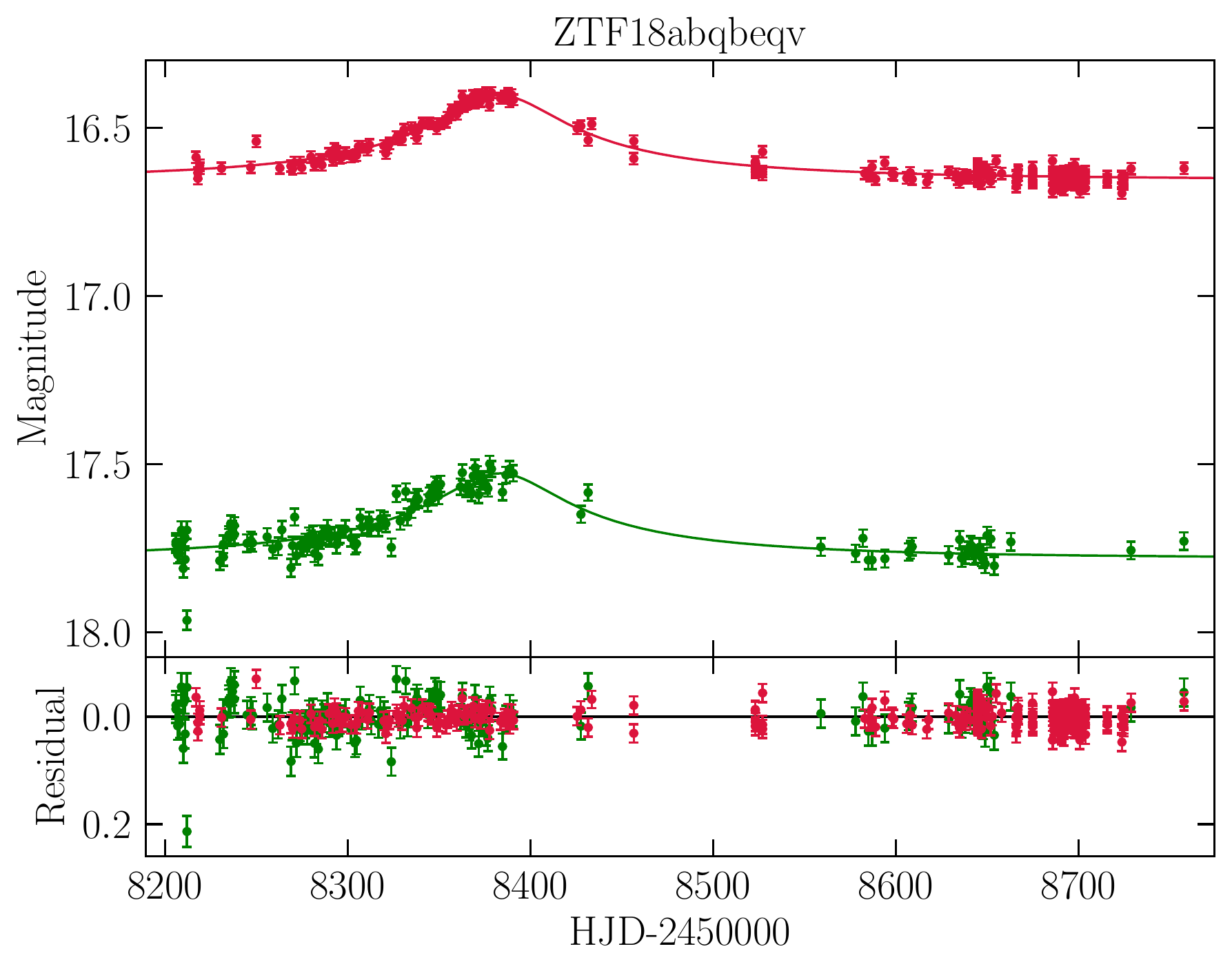}
\includegraphics[width=0.49\textwidth]{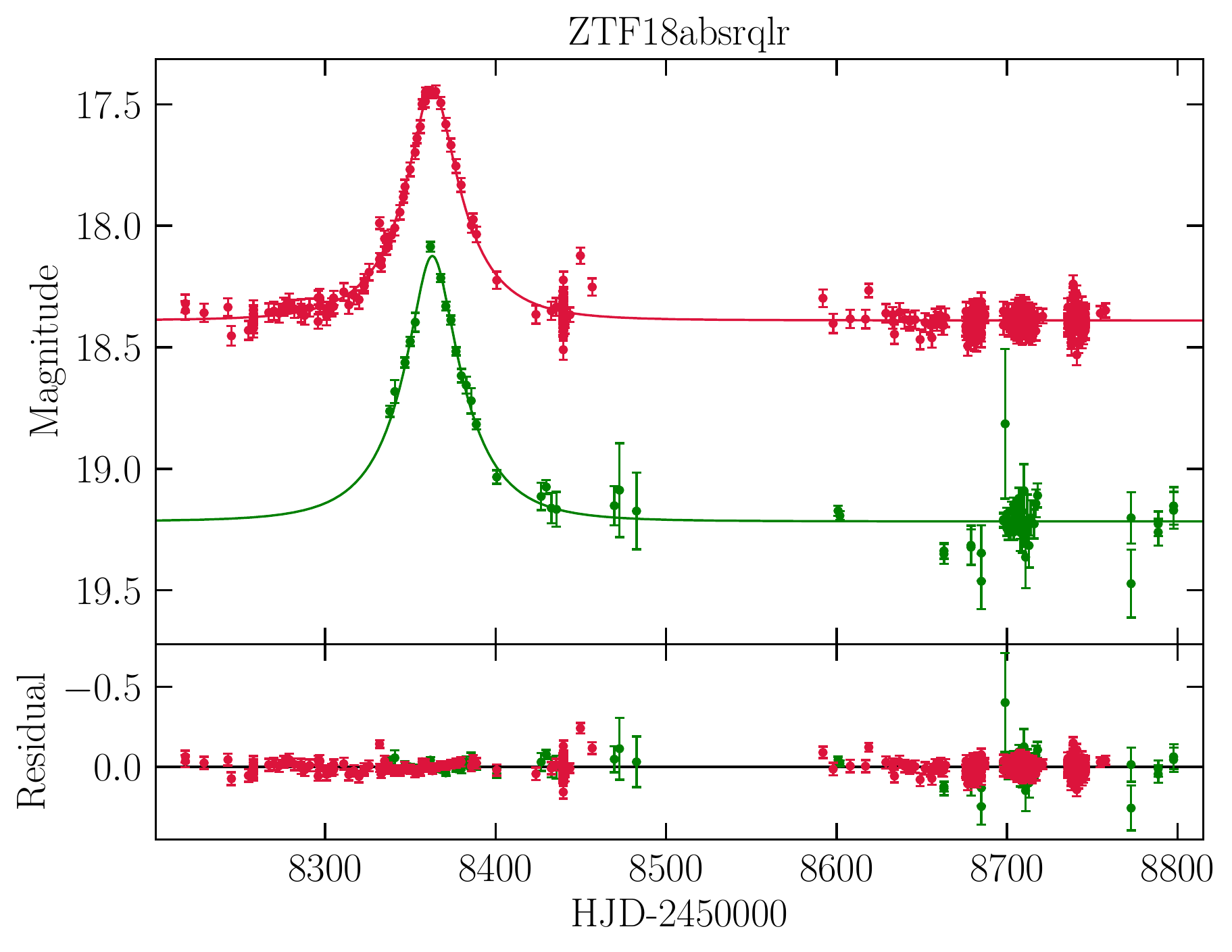}
\includegraphics[width=0.49\textwidth]{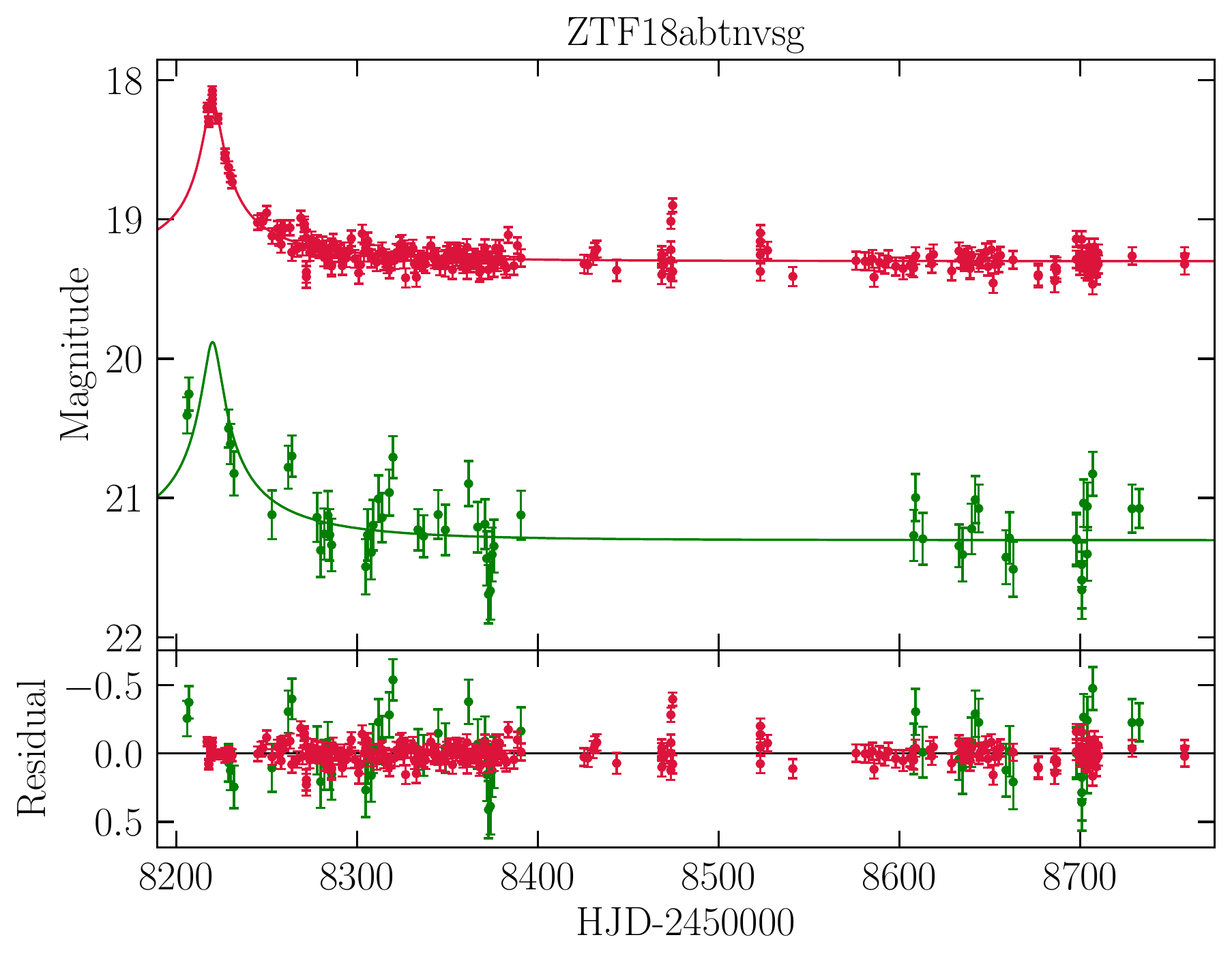}
\includegraphics[width=0.49\textwidth]{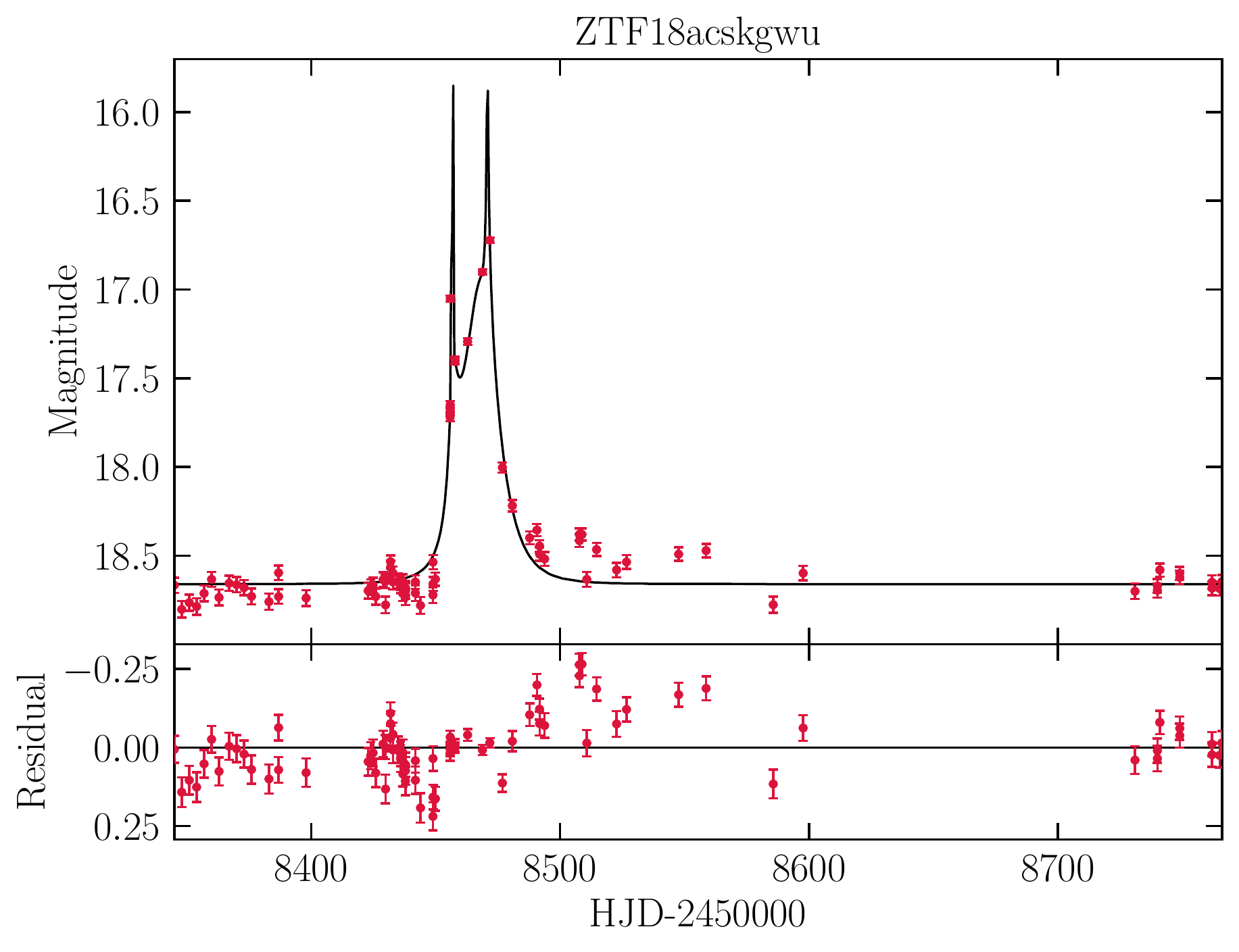}
\includegraphics[width=0.49\textwidth]{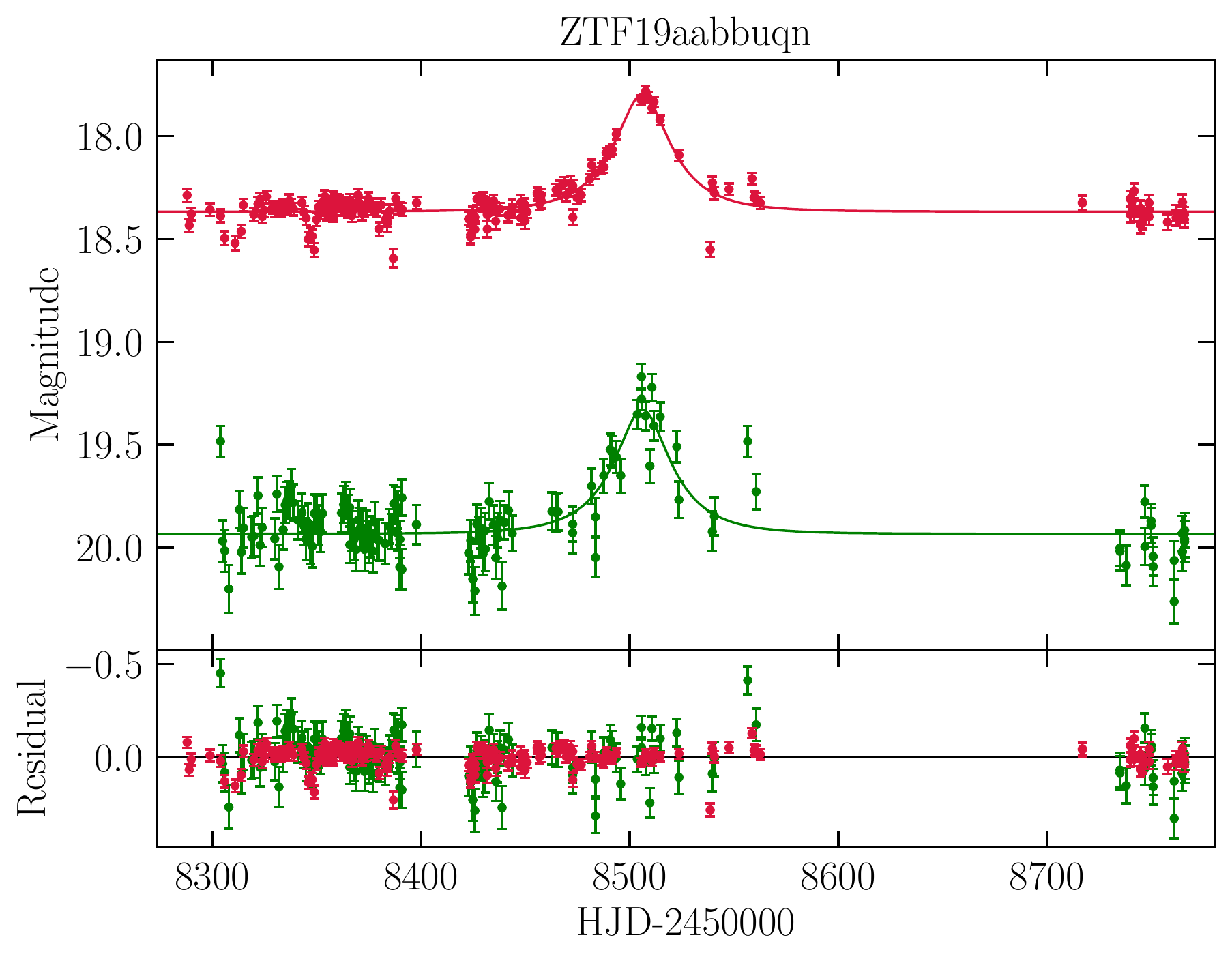}
\caption{Light curves of detected microlensing events. Green and red points were taken in $g$ and $r$ filters, respectively.}
\end{figure}

\begin{figure}
\figurenum{1}
\includegraphics[width=0.49\textwidth]{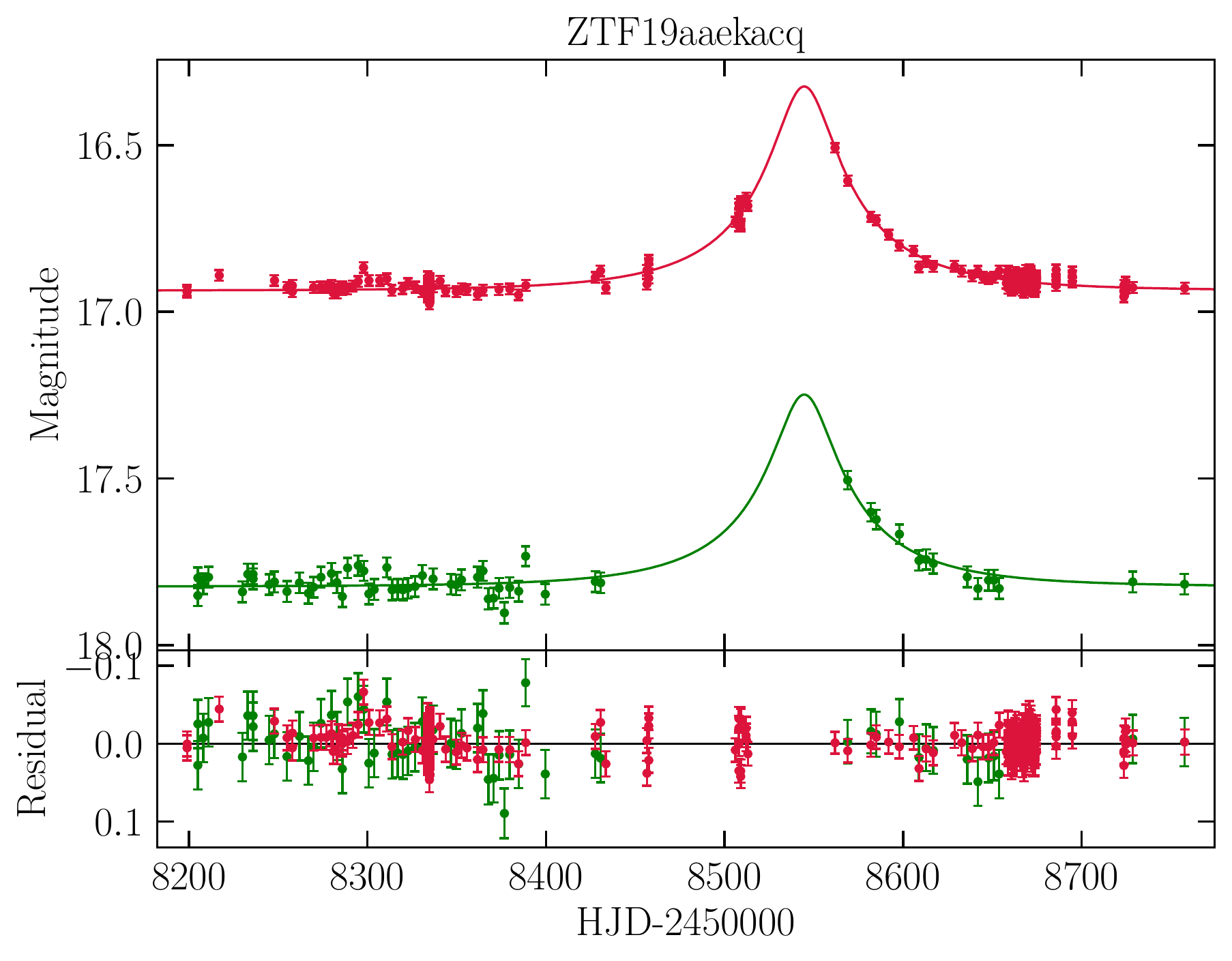}
\includegraphics[width=0.49\textwidth]{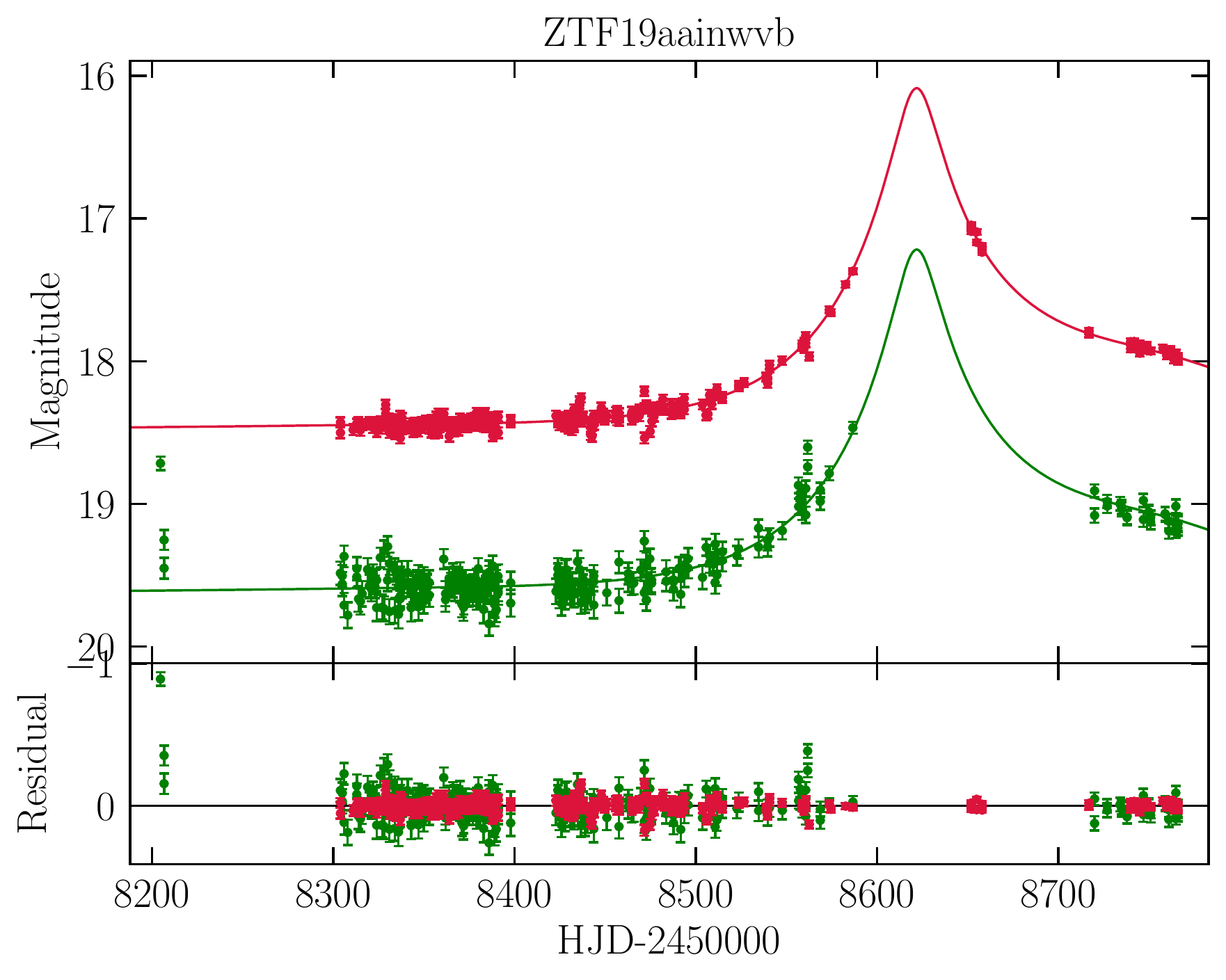}
\includegraphics[width=0.49\textwidth]{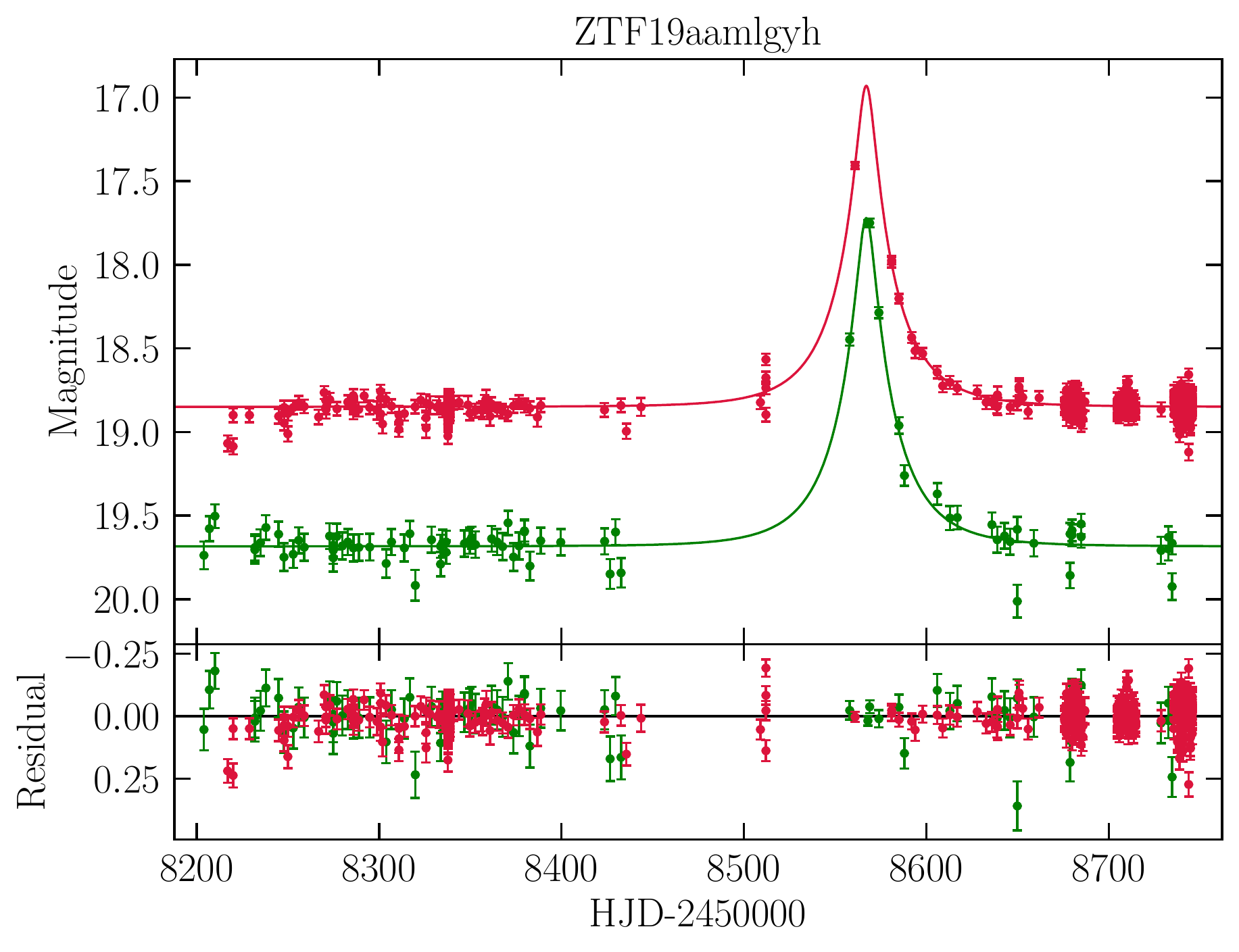}
\includegraphics[width=0.49\textwidth]{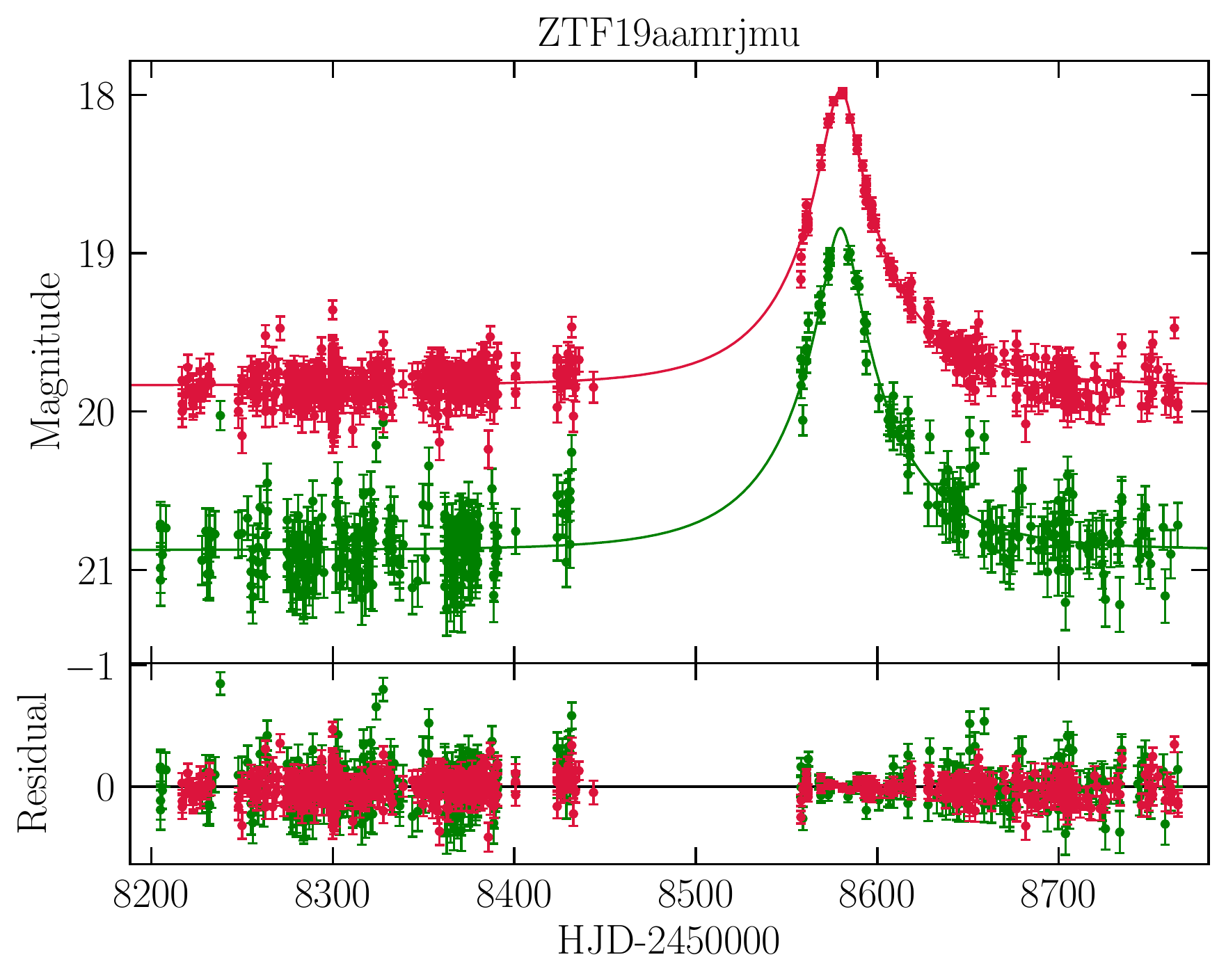}
\includegraphics[width=0.49\textwidth]{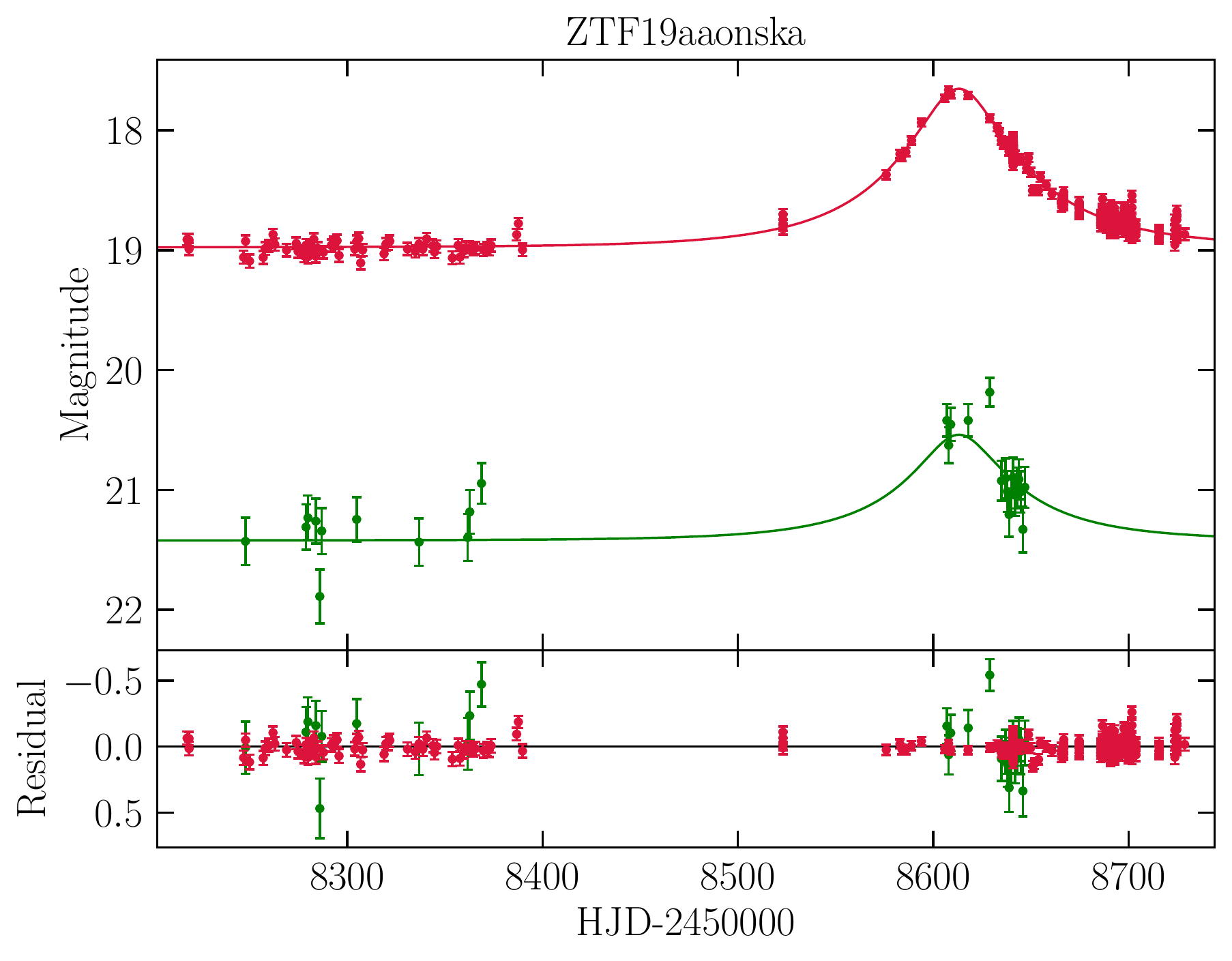}
\includegraphics[width=0.49\textwidth]{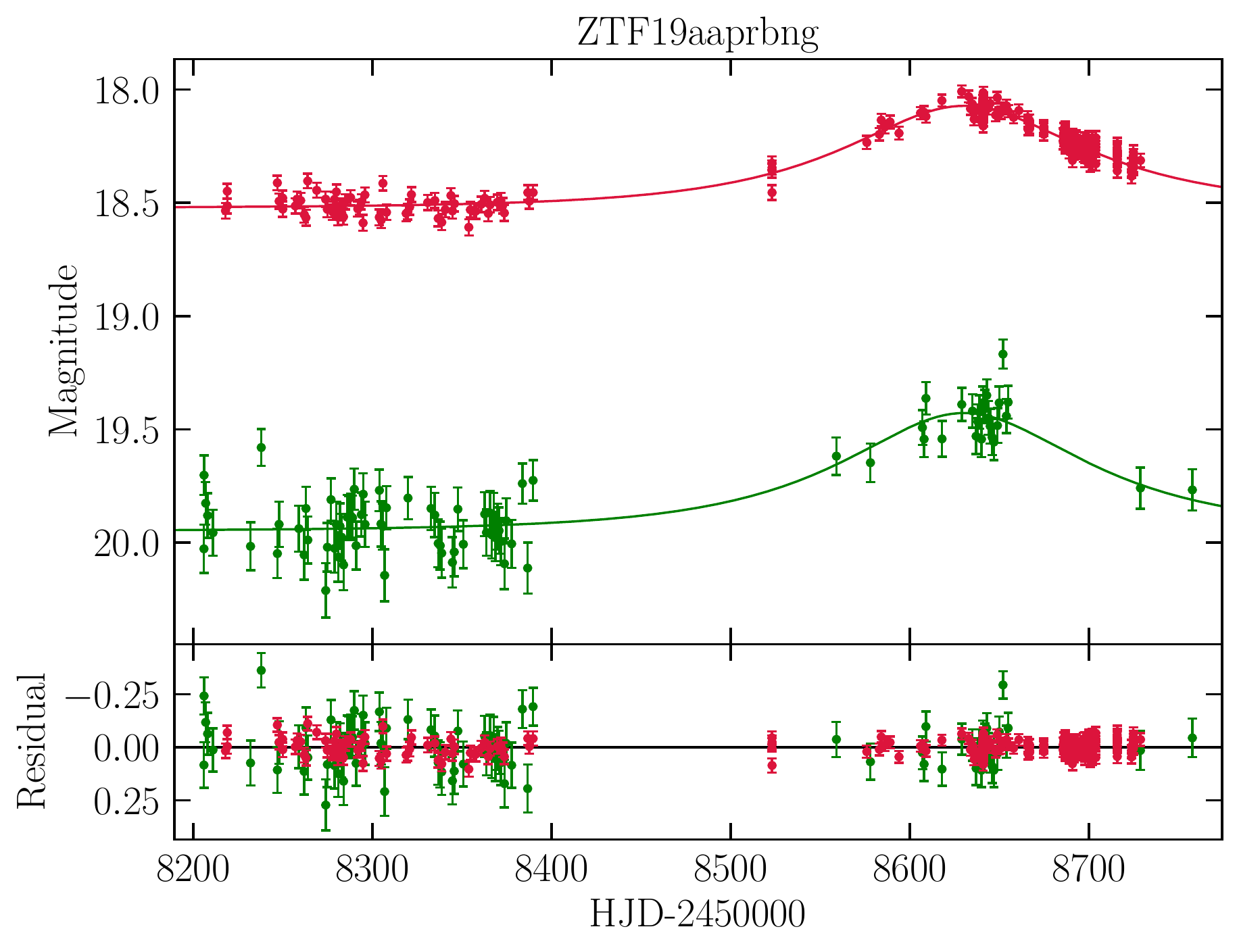}
\caption{Light curves of detected microlensing events. Green and red points were taken in $g$ and $r$ filters, respectively.}
\end{figure}

\begin{figure}
\figurenum{1}
\includegraphics[width=0.49\textwidth]{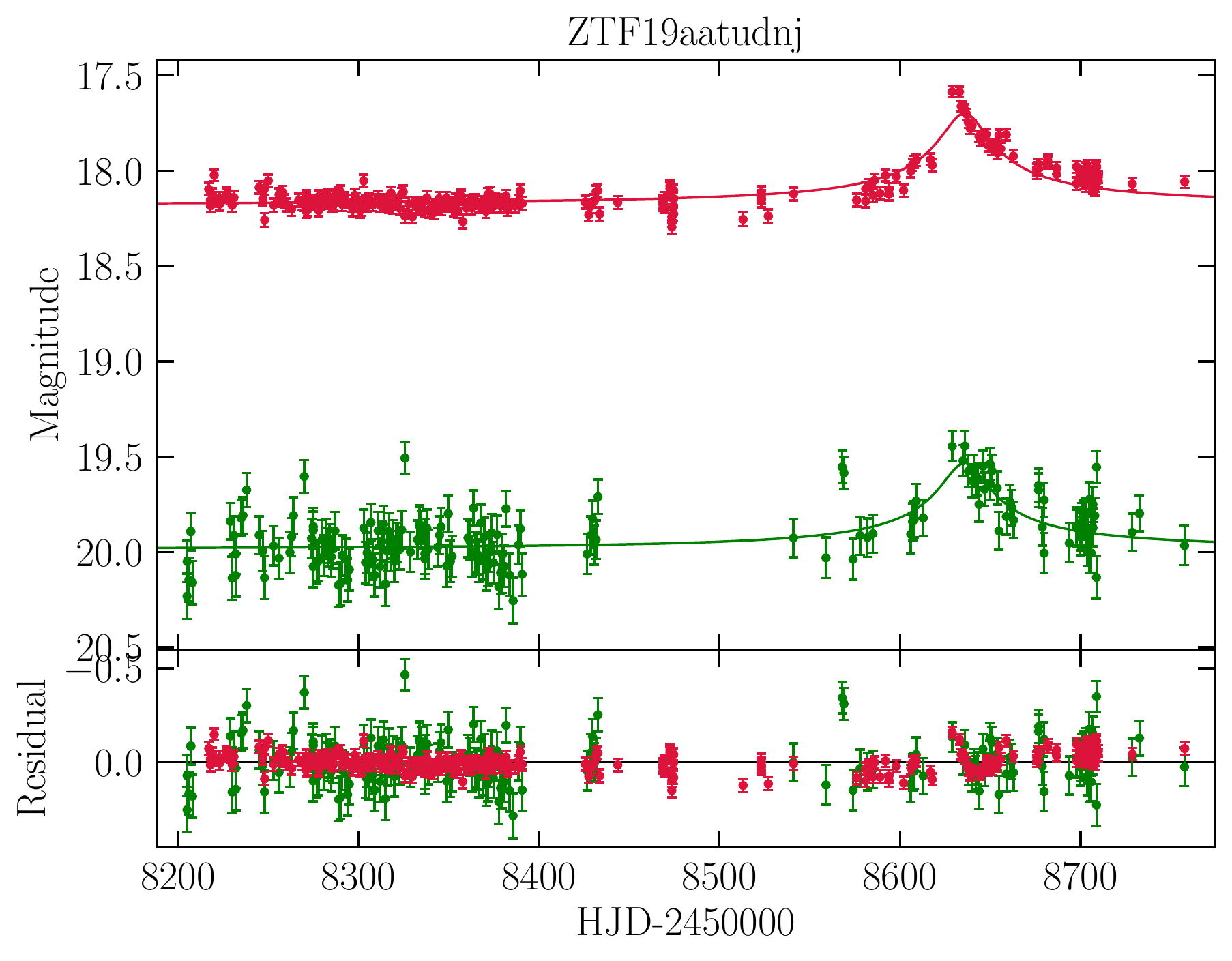}
\includegraphics[width=0.49\textwidth]{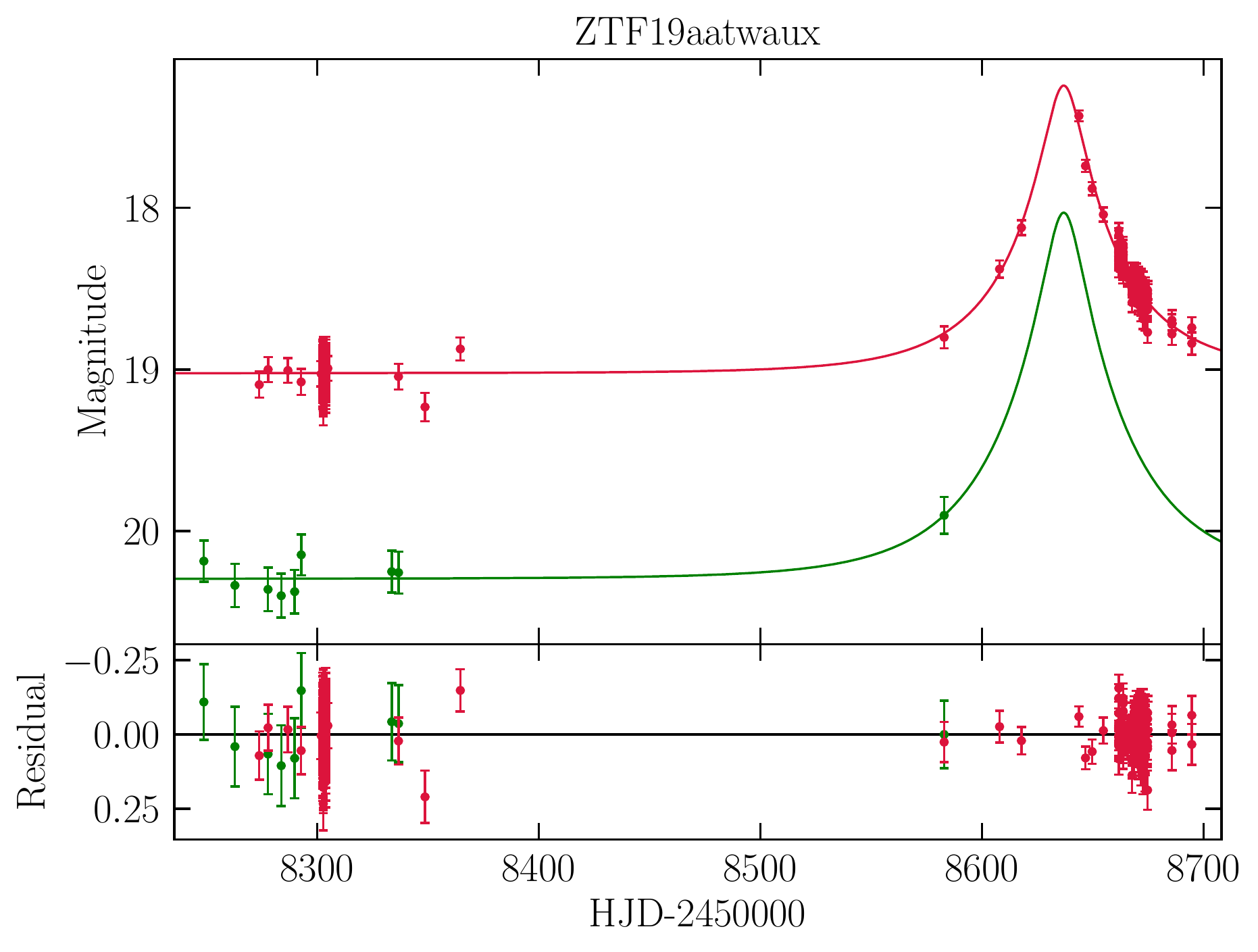}
\includegraphics[width=0.49\textwidth]{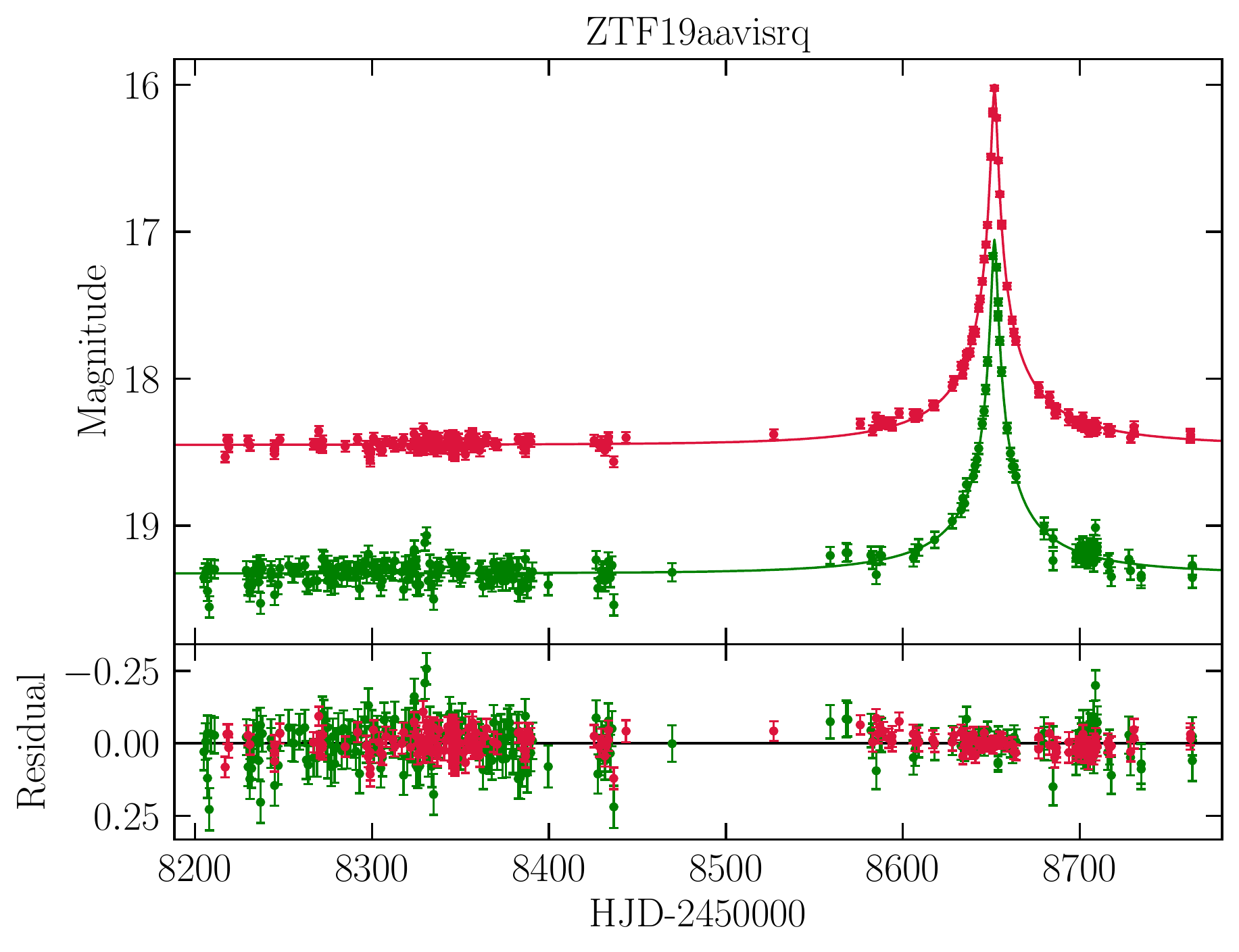}
\includegraphics[width=0.49\textwidth]{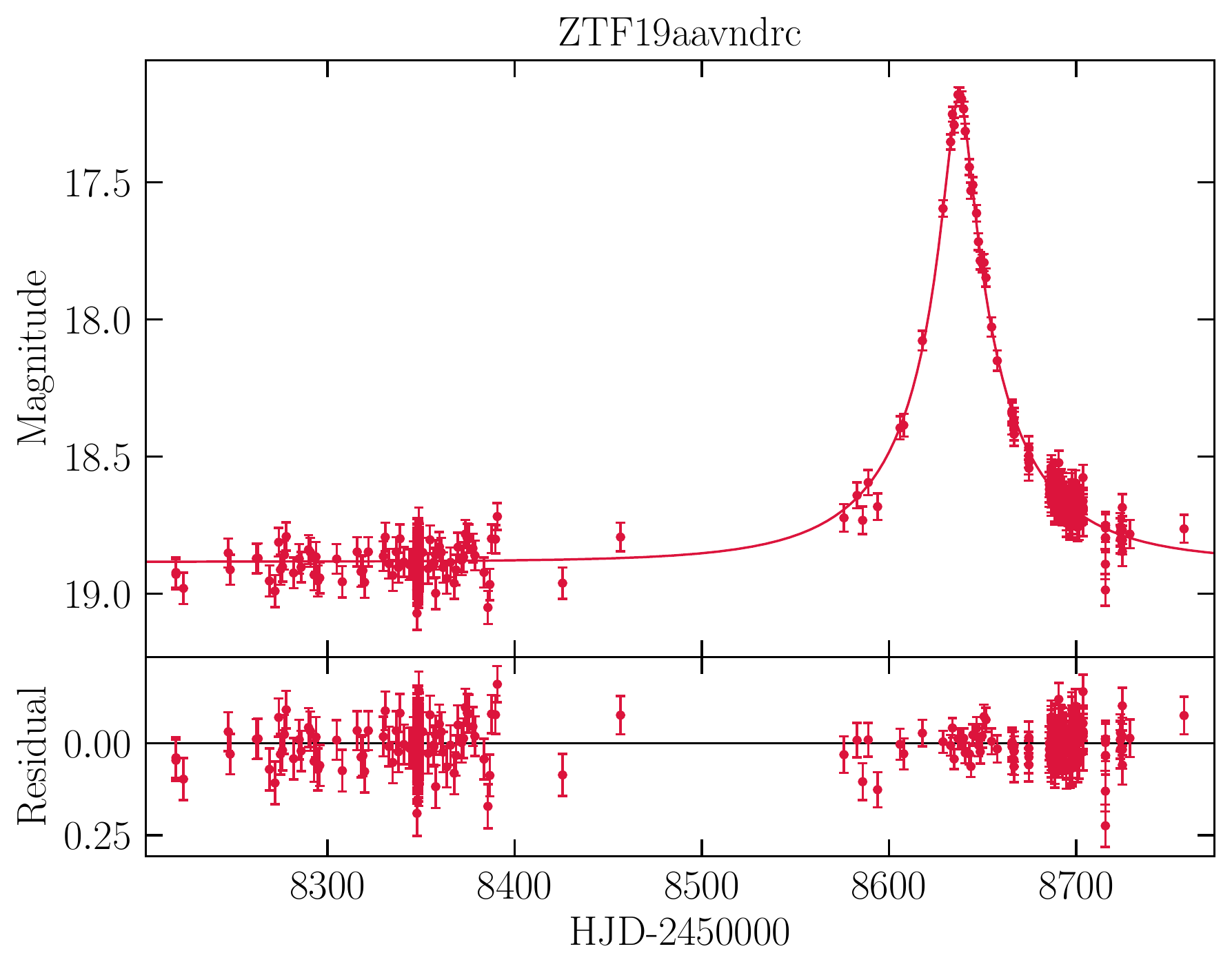}
\includegraphics[width=0.49\textwidth]{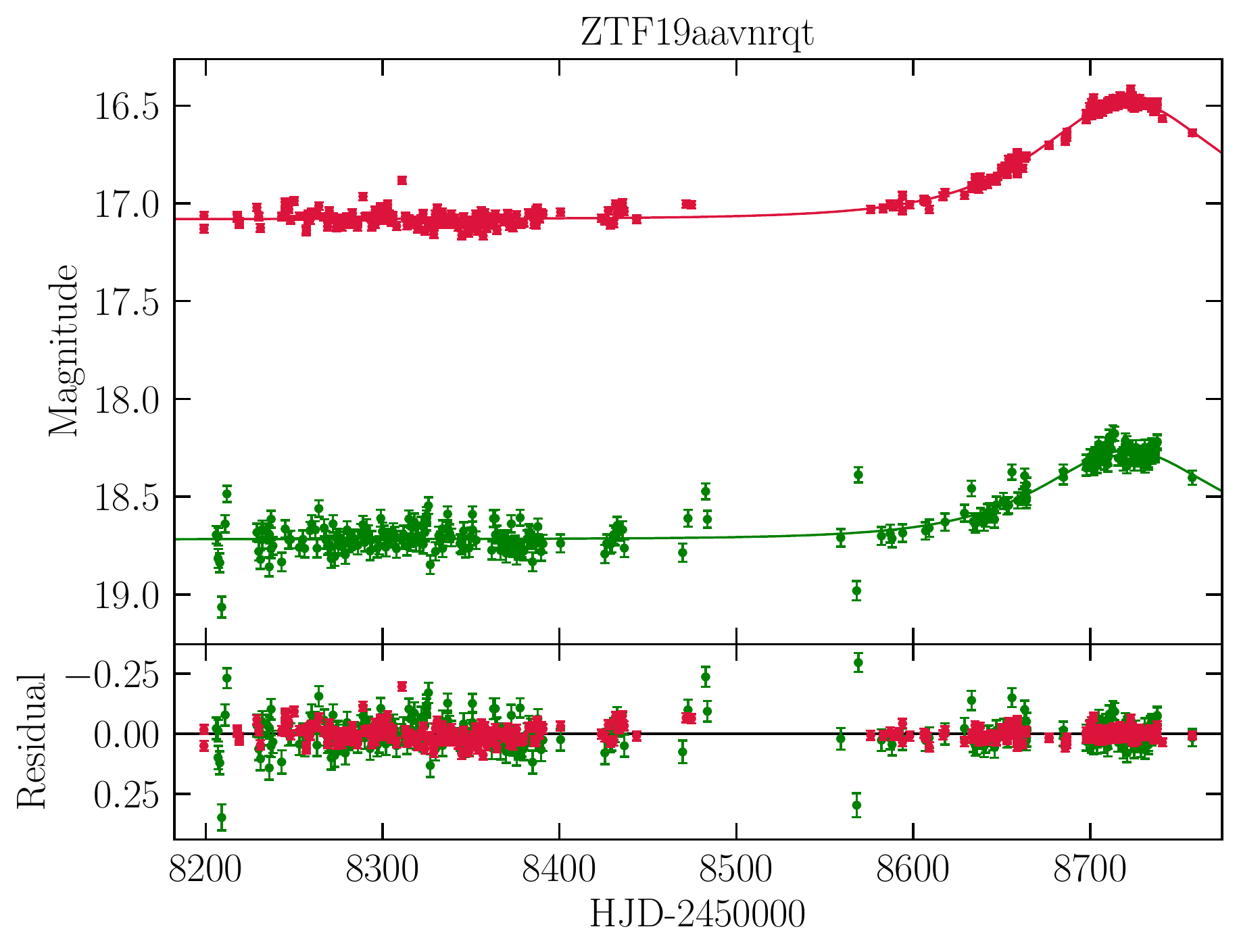}
\includegraphics[width=0.49\textwidth]{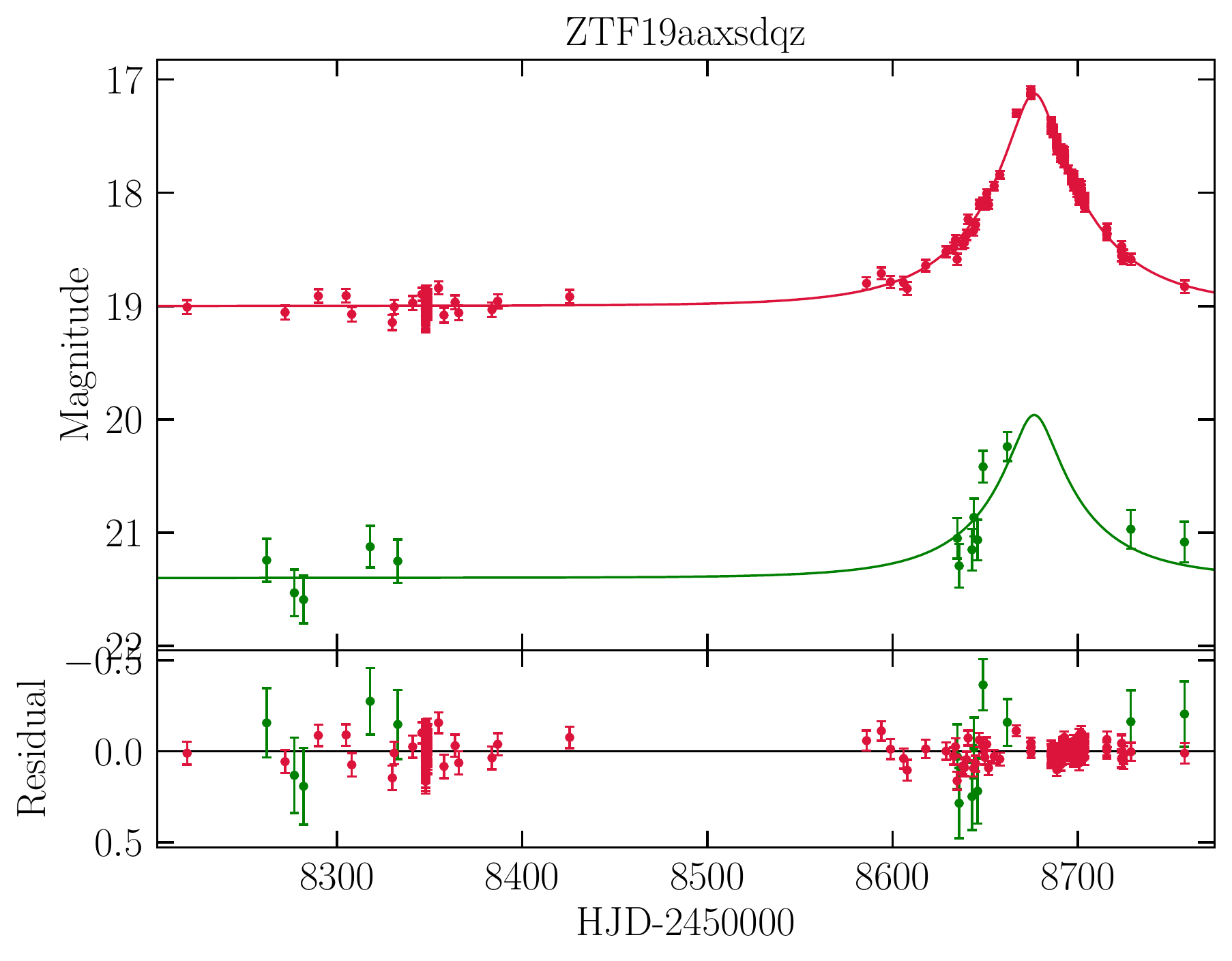}
\caption{Light curves of detected microlensing events. Green and red points were taken in $g$ and $r$ filters, respectively.}
\end{figure}

\end{document}